\documentclass[journal]{IEEEtran}
\usepackage{tikz,cite}
\usepackage{pgfplots}
\usepackage{amsmath}
\usepackage{textcomp}
\usepackage{tikz,setspace}
\usetikzlibrary{backgrounds,intersections}
\usepackage{pgfplots}
\usepackage{comment}
\usepackage{multirow}
\usepackage{amsmath,amssymb}
\usepackage{bm}
\usepackage{bbm}
\usepackage{acronym}
\usepackage[colorlinks=true,bookmarks=false,citecolor=blue,urlcolor=blue]{hyperref}

\newtheorem{example}{Example}

\pgfplotsset{compat=newest}

\colorlet{figyellow}{yellow!40!white}
\colorlet{figred}{red!50!white}
\colorlet{figblue}{blue!20!white}
\colorlet{figgreen}{green!30!white}
\colorlet{figgray}{black!10!white}
\definecolor{ALUColor1}{rgb}{0,0.4470,0.7410}
\definecolor{ALUColor2}{rgb}{0.8500,0.3250,0.0980}
\definecolor{ALUColor3}{rgb}{0.9290,0.6940,0.1250}
\definecolor{ALUColor4}{rgb}{0.4940,0.1840,0.5560}
\definecolor{ALUColor5}{rgb}{0.4660,0.6740,0.1880}
\definecolor{ALUColor6}{rgb}{0.3010,0.7450,0.9330}
\definecolor{ALUColor7}{rgb}{0.6350,0.0780,0.1840}

\acrodef{ADC}{analog-to-digital converter}
\acrodef{AIR}{achievable information rate}
\acrodef{AWGN}{additive white Gaussian noise}
\acrodef{BER}{bit error rate}
\acrodef{BICM}{bit-interleaved coded modulation}
\acrodef{BMD}{bit metric decoder}
\acrodef{BSC}{binary symmetric channel}
\acrodef{CM}{coded modulation}
\acrodef{DAC}{digital-to-analog converter}
\acrodef{DMC}{discrete memoryless channel}
\acrodef{DSP}{digital signal processing}
\acrodef{EDFA}{Erbium-doped fiber amplifier}
\acrodef{FEC}{forward error correction}
\acrodef{GMI}{generalized mutual information}
\acrodef{GPU}{graphical processing unit}
\acrodef{IID}{independent and identically distributed}
\acrodef{KDE}{kernel density estimator}
\acrodef{LLR}{log-likelihood ratio}
\acrodef{LDPC}{low-density parity-check}
\acrodef{MI}{mutual information}
\acrodef{ML}{maximum likelihood}
\acrodef{MLC}{multi-level coding}
\acrodef{MSD}{multi-stage decoding}
\acrodef{NB}{nonbinary}
\acrodef{PDF}{probability density function}
\acrodef{QAM}{quadrature amplitude modulation}
\acrodef{RS}{Reed-Solomon}
\acrodef{SER}{symbol error rate}
\acrodef{SNR}{signal-to-noise ratio}
\acrodef{SSMF}{standard single-mode fiber}
\acrodef{WDM}{wavelength division multiplex}
\acrodef{WSS}{wavelength selective switch}

\newcommand{\Xs}{{X}}
\newcommand{\Us}{{U}}
\newcommand{\Ys}{{Y}}
\newcommand{\Zs}{{Z}}

\def\doubleunderline#1{\underline{\underline{#1}}}

\begin{document}
\title{Performance Prediction of Nonbinary Forward Error Correction in Optical Transmission Experiments}

\author{Laurent Schmalen, Alex Alvarado, and Rafael Rios-M\"{u}ller
\thanks{Parts of this paper have been presented at the 2016 Optical Fiber Communication Conference (OFC), Anaheim, CA, USA, Mar. 2016 in paper M2A.2~\cite{schmalen_ofc16}.}
 \thanks{L. Schmalen is with Nokia Bell Labs, Stuttgart, Germany (e-mail: first.last@nokia-bell-labs.com).}
\thanks{A. Alvarado is with the Optical Networks Group, Dept. of Electronic \& Electrical Engineering, University College London (UCL), London, WC1E 7JE, UK.}
\thanks{R. Rios-M\"{u}ller is with Nokia Bell Labs, Villarceaux, France.}
\thanks{L. Schmalen was financially supported by the CELTIC EUREKA project SENDATE-TANDEM (Project ID C2015/3-2) which is partly funded by the German BMBF (Project ID 16KIS0450K). A. Alvarado was supported by the Engineering and Physical Sciences Research Council (EPSRC) project UNLOC (EP/J017582/1), UK.}
}

\markboth{Preprint, \today}{}

\IEEEspecialpapernotice{(Invited Paper)}

\maketitle

\begin{abstract}
In this paper, we compare different metrics to predict the error rate of optical systems based on nonbinary forward error correction (FEC). It is shown that an accurate metric to predict the performance of coded modulation based on nonbinary FEC is the mutual information. The accuracy of the prediction is verified in a detailed example with multiple constellation formats and FEC overheads, in both simulations and optical transmission experiments over a recirculating loop.  It is shown that the employed FEC codes must be universal if performance prediction based on thresholds is used. A tutorial introduction into the computation of the thresholds from optical transmission measurements is also given.
\end{abstract}


\section{Introduction and Motivation}

\IEEEPARstart{M}{any} optical transmission experiments do not include \ac{FEC}. The reasons for this are that often, \ac{FEC} development is still ongoing, or \ac{FEC} developers are physically remote from the experiment. Often, researchers would also like to reuse experimental data obtained in expensive optical transmission experiments  to evaluate the performance of different \ac{FEC} schemes, without needing to redo the transmission experiment and/or signal processing.  Therefore, \emph{thresholds} are commonly used to decide whether the \ac{BER} after \ac{FEC} decoding is below the required target BER, which can be in the range of $10^{-13}$ to $10^{-15}$. The most commonly used threshold in the optical communications literature is the pre-FEC~\ac{BER}. 

The use of thresholds is also very convenient in practice because very low post-FEC BER values are hard to estimate. The conventional design strategy has therefore been to experimentally demonstrate (or simulate) systems without FEC encoding and decoding, and to optimize the design for a much higher BER value, the so-called ``FEC limit'' or ``FEC threshold''. This approach relies on the strong assumption that a certain BER without coding can be reduced to the desired post-FEC BER by previously verified FEC implementations, regardless of the system under consideration. 

\begin{figure*}[t]
\centerline{
\includegraphics{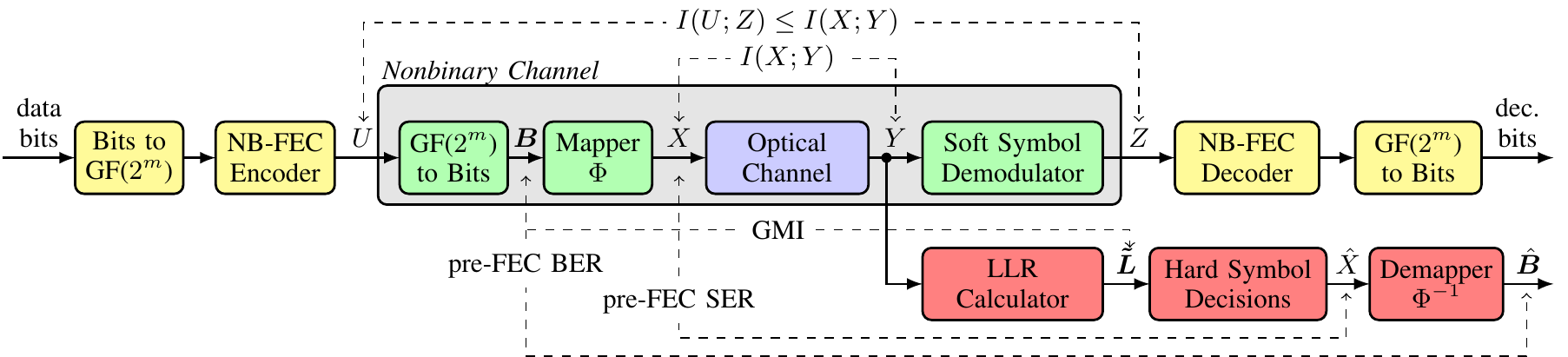}}
\caption{System model of optical transmission based on \acs{NB}-\acs{CM} and the measurement of various system parameters.}
\label{fig:system_model}
\end{figure*}

Using pre-FEC BER thresholds is very popular in the literature and has been used for example in the record experiments based on $2048$ \ac{QAM} for single-core \cite{Beppu15} and multi-core \cite{Qian13ecoc} fibers. This threshold indeed gives accurate post-FEC BER predictions if three conditions are satisfied. First, bit-level interleaving must be used to guarantee independent bit errors. Second, the \ac{FEC} under consideration must be binary and universal, and lastly, the decoder is based on hard decisions (bits) rather than soft decisions. Recently, however, it was shown in~\cite{Alvarado2015b_JLT,Alvarado16a} that the pre-FEC~\ac{BER} fails at predicting the post-\ac{FEC}~\ac{BER} of binary \emph{soft-decision} \ac{FEC}. This was shown for both turbo codes and \ac{LDPC} codes, in the linear and nonlinear regimes, and in both simulations and optical experiments. Furthermore, \cite{Alvarado2015b_JLT} also showed that a better predictor in this case is the \ac{GMI}\footnote{Also known as the BICM capacity or parallel decoding capacity.} \cite[Sec.~3]{Fabregas08_Book}, \cite[Sec.~4.3]{Alvarado15_Book}, \cite{Bulow2011b,Alvarado2015_JLT} and suggested to replace the pre-FEC~\ac{BER} threshold by a ``GMI threshold''.

The rationale for using the \ac{GMI} as a metric to characterize the performance of binary soft-decision \ac{FEC} is that the \ac{GMI} is an \ac{AIR} for \ac{BICM}\cite{Fabregas08_Book,Alvarado15_Book}, often employed as a pragmatic approach to \acf{CM}. For square \ac{QAM} constellations, \ac{BICM} operates close to capacity with moderate effort, and thus, it is an attractive CM alternative. However, for most nonsquare \ac{QAM} constellations, \ac{BICM} results in unavoidable performance penalties. For these modulation formats, other \ac{CM} schemes such as \ac{NB} \ac{FEC}~\cite{DjordjevicNB} and multi-level coding with multi-stage decoding~\cite{BeygiCM} can be advantageous. Furthermore, \ac{BICM} is not expected to be the most complexity-efficient coded modulation scheme for short reach and metro optical communications with higher order modulation. The reason is that the \ac{DSP} implementation needs to work at the transmission baud rate, but the \ac{FEC} decoder needs to operate at $m$ times the \ac{DSP} rate, if $2^m$-ary higher order modulation formats are used. For these applications, multi-level coding~\cite{wachsmann1999multilevel,Bisplinghoff:16} or \ac{NB}-\ac{FEC} may be good candidates and for these, the throughput is in the same order as for the \ac{DSP}. Although most nonbinary \ac{FEC} schemes are considerably more complex to implement than their binary counterparts, recent advances~\cite{montorsi2012analog},\cite{awais2014vlsi} show that very low-complexity nonbinary FEC schemes for higher order constellations can be implemented  using for instance the numerically stable algorithm presented in~\cite{beermann2015gpu}.

In this paper, we investigate the performance prediction of \ac{NB} soft-decision \ac{FEC} (NB-FEC) and show that an accurate threshold in this case is the \ac{MI}\cite{Shannon48}. The \ac{MI} was previously introduced in~\cite{LevenMI} to assess the performance of differentially encoded quaternary phase shift keying and was shown to be a better performance indicator than the pre-FEC BER. The use of MI as a post-FEC BER predictor for capacity-approaching nonbinary FEC was also conjectured in \cite[Sec.~V]{Alvarado2015b_JLT} and was previously suggested in \cite{Brueninghaus05,Wan06} in the context of wireless communications.

The main contribution of this paper is to show that the \ac{MI} is an accurate threshold for a \ac{CM} scheme based on \acs{NB} \ac{LDPC} codes. This is verified in both an \ac{AWGN} simulation and in two optical experiments using 8-QAM constellations. We show that the \ac{MI} allows us to accurately predict the post-FEC performance of \ac{NB} \ac{LDPC} schemes and also show that other commonly used thresholds (such as pre-FEC BER, pre-FEC \ac{SER} and bit-wise \ac{GMI}) fail in this scenario.

This paper is organized as follows. In Sec.~\ref{sec:system} we describe the system model we use and lay down some information theory preliminaries. Afterwards, in Sec.~\ref{sec:thresholds} we show what thresholds we should use to predict the performance of \ac{NB} \ac{FEC} schemes. In Sec.~\ref{sec:experiment}, we verify our predictors with a simulation example, a back-to-back experiment and a transmission experiment over a recirculating loop. Finally, in Sec.~\ref{sec:universalityrevisit}, we discuss code universality and give guidelines for using the proposed thresholds.

\begin{figure*}[t]
\centerline{
\includegraphics{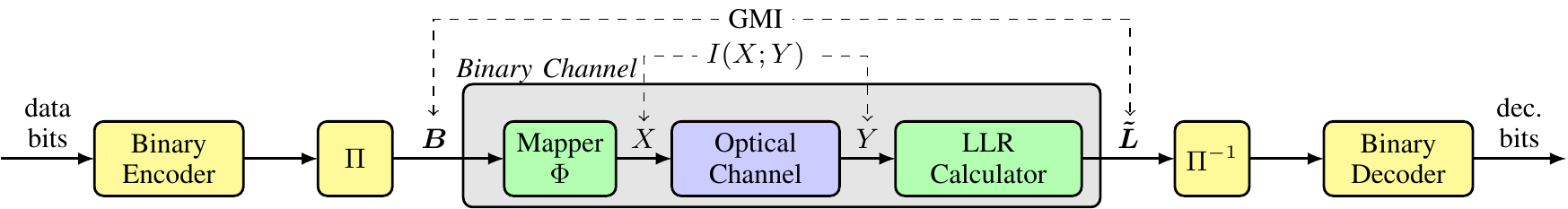}}
\caption{System model of optical transmission based on BICM.}
\label{fig:system_model_bicm}
\end{figure*}

\section{System Model and Preliminaries}\label{sec:system}

\subsection{System Model}

Fig.~\ref{fig:system_model} shows the \acs{NB}-\ac{CM} scheme under consideration. The data bits are mapped to \acs{NB} symbols from GF($2^m$) using a one-to-one (i.e., invertible) mapping function, then encoded by an {\acs{NB}-\ac{FEC}} with rate $R$, and then mapped to $D$-dimensional constellation symbols from the set $\mathcal{S}:=\{s_1,\ldots,s_M\}$, where $|\mathcal{S}|=2^{m}=M$ and $s_i\in\mathbb{R}^D$. Frequently, $D=2$ (with complex symbols), but in optical communications, also $D=4$~\cite{Agrell2009_JLT,Bulow2011b,Alvarado2015_JLT,Eriksson_MI_JLT} and $D=8$ \cite{ErikssonOFC14,Koike-AkinoECOC13} are used. As will become obvious later, the mapping to symbols is shown in two stages in Fig.~\ref{fig:system_model}, namely first mapping the \acs{NB} symbols $\Us\in\left\{1,2,\ldots,M\right\}$ to bit patterns $\boldsymbol{B}$ of $m$ bits, and mapping these to constellations symbols $X\in\mathcal{S}$. In some cases, we require the combination of bit mapper and mapper $\Phi$, which we denote by $\phi(i) = s_i$ and which maps an integer $i$ to a modulation symbol $s_i$.

The constellation symbol $s_i\in\mathcal{S}$ is transmitted with \emph{a priori} probability $P(\Xs=s_i) := \lambda_{i}$ through an ``optical channel''\footnote{We use upper-case letters (e.g., $X$) to denote random variables and lower-case letters (e.g., $x$) to denote realizations of this random variable. We use boldface upper-case letters (e.g., $\bm{X}$) to denote sequences of random variables and boldface lower-case letters (e.g., $\bm{x}$) for their realizations. Sets are denoted by calligraphic letters (e.g., $\mathcal{S}$). $\lVert\cdot\rVert$ is used to denote the $L_2$ norm.}. Most communication systems transmit equiprobable symbols, i.e., $\lambda_{i} = 1/M$, $\forall i$. However, in the case of probabilistic shaping~\cite{buchali_jlt15,bocherer_bandwidth_2015,FehenbergerPTL16}, the probabilities of occurrence of the symbols may differ.
The optical channel\footnote{Also referred to as ``discrete-time (noisy) channel'' in \cite{Agrell16RS}.} takes a sequence of $N_m$ constellation symbols $\bm{x}_1^{N_m} := (x[1], x[2], \ldots, x[N_m])$ and maps them to a waveform $w(t)$ by means of a pulse shaping function $\rho(t)$ with
\[
w(t) = \sum_{\kappa=1}^{N_m}x[\kappa]\cdot \rho(t-\kappa T_s)
\]
with $T_s$ being the symbol period and $\kappa$ the discrete-time index. The optical channel further includes \acp{DAC}, filtering, transmission including amplification, \acp{ADC}, and \ac{DSP} to remove effects of chromatic dispersion, polarization mode dispersion, polarization rotation, phase noise, frequency offset, etc. It further includes matched filtering, equalization and possibly (de-)interleaving.

At the receiver, for each sampled symbol $y[\kappa]$, the soft symbol demodulator (see Fig.~\ref{fig:system_model}) computes $M$ scaled likelihoods (which are proportional to the \emph{a posteriori} probabilities) $q_{\Ys|\Xs}(y|s_i)\lambda_{i}$, where $q_{\Ys|\Xs}(y|s_i)$ is a function that depends on the received $D$-dimensional sampled symbol $y$ and the constellation symbol $s_i\in\mathcal{S}$.  These scaled likelihoods are passed to an {\acs{NB}-\ac{FEC}} decoder. Note that usually, for numerical reasons, a vector of $M-1$ nonbinary \acp{LLR} is computed for each $D$-dimensional received symbol $y$ instead. These (nonbinary) \acp{LLR}\footnote{Strictly speaking, the quantities in~\eqref{eq:llrcomp} are log-\emph{a posteriori} probability (log-APP) ratios. However, in the \ac{FEC} literature, the log-APP ratios are typically also called LLRs, which is why we follow this latter convention here (see also~\cite[Sec.~4.5.3]{ryan2009channel} and~\cite[p.~58]{Alvarado15_Book}).} are given by
\begin{align}
L_i(y) = \ln\left(\frac{q_{\Ys|\Xs}(y|s_i)}{q_{\Ys|\Xs}(y|s_1)}\right) + \ln\left(\frac{\lambda_{i}}{\lambda_{1}}\right),\ \forall i\in\{2,\ldots, M\}.\label{eq:llrcomp}
\end{align}
Ideally, the receiver knows the (averaged) optical channel transition \ac{PDF} $p_{\Ys|\Xs}(y|s_i)$, applies sufficiently long interleaving, and sets $q_{\Ys|\Xs}(y|x) = p_{\Ys|\Xs}(y|x)$ in \eqref{eq:llrcomp}. Usually, however, the exact channel transition \ac{PDF} is not known at the receiver, or the computation of the \acp{LLR} is too involved using the true \ac{PDF}, which is why often approximations are used. In this case $q_{\Ys|\Xs}(y|x) \neq p_{\Ys|\Xs}(y|x)$, and thus, we say that the receiver is \emph{mismatched}~\cite{merhav1994information}. Often, for instance, the (multivariate) Gaussian \ac{PDF} is assumed at the receiver, i.e., $q_{\Ys|\Xs}(y|s_i) = q_{\textsf{awgn}}(y|s_i)$, where
\[
q_{\textsf{awgn}}(y|s_i) := \frac{\exp\left(-\frac{1}{2}(y-s_i)^T\bm{\Sigma}^{-1}(y-s_i)\right)}{\sqrt{(2\pi)^D|\bm{\Sigma}|}}.
\]
In~\cite{Eriksson_MI_JLT}, different approximations are compared for $D=4$  and it was found that the circularly symmetric Gaussian approximation with diagonal covariance matrix $\bm{\Sigma}$ reliably approximates the true \ac{PDF} unless the input power is increased to very high levels. Besides, the Gaussian \ac{PDF} has also been shown to be a good approximation for the true \ac{PDF} in case of uncompensated fiber links with coherent reception~\cite{poggiolini2012gn}. Furthermore, using a Gaussian \ac{PDF} also simplifies the numerical computation of the \acp{LLR}. 

A predominant case is $D=2$ (e.g., QAM constellations detected independently in each polarization) with circularly symmetric noise (diagonal $\bm{\Sigma}$) and variance $\sigma_n^2$ per dimension. This is the case on which we focus on this paper and which is also dominant in coherent long-haul dispersion uncompensated links~\cite{Eriksson_MI_JLT}. In this case
\begin{align}\label{q.awgn}
q_{\textsf{awgn}}(y|s_i)\big|_{D=2} = \frac{1}{2\pi\sigma_n^2}\exp\left(-\frac{\left\lVert y-s_i\right\rVert^2}{2\sigma_n^2}\right)\,.
\end{align}
Assuming equally likely symbols ($\lambda_i = 1/M$), the \acp{LLR} in \eqref{eq:llrcomp} are given by 
\begin{align}\label{eq:llrcomp:2}
L_i(y) = \frac{1}{2\sigma_n^2}(\left\lVert y-s_1\right\rVert^2 - \left\lVert y-s_i\right\rVert^2). 
\end{align}
After \ac{LLR} computation, the \ac{NB} soft-decision \ac{FEC} decoder (e.g., a nonbinary LDPC decoder) takes these \acp{LLR} and estimates the transmitted \acs{NB} symbols, which are later converted into decoded bits. Here we only assume that the nonbinary \ac{FEC} is matched to the constellation, i.e., each nonbinary symbol of the \ac{FEC} code can be mapped to $m=\log_2(M)$ bits. This allows us to consider nonbinary \ac{LDPC} codes defined over either the Galois field $\text{GF}(2^m)$ or the ring $\mathbb{Z}_M$ of integers modulo $M$. We further assume that \emph{soft decision} decoding is carried out, see, e.g.,~\cite{beermann2015gpu}. For other, low complexity versions of that algorithm, we refer the interested reader to the references in~\cite{beermann2015gpu}.

\subsection{Bit-Interleaved Coded Modulation (BICM)}\label{sec:bicm}
In optical communications, often the pragmatic \ac{BICM} scheme is used. The system model of \ac{BICM} is shown in Fig.~\ref{fig:system_model_bicm}. We only describe a simplified version here. For more details, we refer the interested reader to~\cite{Alvarado15_Book,BeygiCM} and references therein. \ac{BICM} is based on a binary \ac{FEC} code. The binary output of the \ac{FEC} encoder is interleaved\footnote{Often, the interleaver is considered to be part of the \ac{FEC} encoder, for instance if random \ac{LDPC} codes are used} by a permutation function $\Pi$. The resulting interleaved bit stream $\boldsymbol{B}$ is then mapped to modulation symbols $\Xs$ using the mapper $\Phi$ described above.

At the receiver, we use a \ac{BMD} to compute \acp{LLR} $\boldsymbol{\tilde{L}}$ for the individual bits of the bit stream $\boldsymbol{B}$. In Fig.~\ref{fig:system_model_bicm}, the \ac{BMD} is denoted \emph{LLR Calculator}. The  \acp{LLR} computed by the \ac{BMD} are given by
\begin{align}
\tilde{L}_i(y) = \log\left(\frac{\sum_{s\in\mathcal{S}_{0,i}}q_{\Ys|\Xs}(y|s)\lambda_{\phi^{-1}(s)}}{\sum_{s\in\mathcal{S}_{1,i}}q_{\Ys|\Xs}(y|s)\lambda_{\phi^{-1}(s)}}\right),\ \forall i\in\{1,\ldots,m\}\label{eq:llr_bicm}
\end{align}
where $\mathcal{S}_{b,i}$ is the set of constellation symbols where the $i$-bit of the binary label takes on the value $b$. In the practically dominant case with equiprobable symbols ($\lambda_i = 1/M$), we get
\begin{align*}
\tilde{L}_i(y) = \log\left(\frac{\sum_{s\in\mathcal{S}_{0,i}}q_{\Ys|\Xs}(y|s)}{\sum_{s\in\mathcal{S}_{1,i}}q_{\Ys|\Xs}(y|s)}\right)
\end{align*}
The stream of \acp{LLR} $\boldsymbol{\tilde{L}}$ is then de-interleaved by the inverse permutation $\Pi^{-1}$ and then fed to a conventional soft-decision binary \ac{FEC} decoder.

The comparison of \eqref{eq:llr_bicm} with \eqref{eq:llrcomp} clearly shows the difference between nonbinary \ac{CM} and \ac{BICM}. In the nonbinary case, we compute a vector of \acp{LLR} containing $M-1$ values for each channel observation $\Ys$. In contrast, for \ac{BICM}, we only compute $m=\log_2(M)$ \acp{LLR} per channel observation. Clearly, there is a compression of information which is available for the \ac{FEC} decoder. Fascinatingly, the loss of information from this compression can be made negligible in many practical cases, e.g., with square \ac{QAM} constellations and Gray mapping~\cite{Alvarado15_Book,BeygiCM}. The loss of information may however become important for other constellations.

\subsection{FEC Universality}\label{sec:universality}

When assessing and comparing the performance of different modulation formats and different transmission scenarios (e.g., fiber types, modulators, converters, etc.) based on \emph{thresholds}, it is important to understand the concept of \ac{FEC} \emph{universality}. A pair of \ac{FEC} code and its decoder are said to be \emph{universal} if the performance of the code (measured in terms of post-FEC \ac{BER} or \ac{SER}) does not depend on the nonbinary channel (with input $U$ and output $Z$ when referring to Fig.~\ref{fig:system_model}), provided that the channel has a fixed mutual information $I(U;Z)$.

When we refer to ``the channel'', we consider the whole transmission chain between the \ac{NB}-\ac{FEC} encoder output $\Us$ and the decoder input $\Zs$ including modulation and demodulation, \ac{DSP}, \acp{ADC} and \acp{DAC}, optical transmission and amplification including noise. We say that the channel changes if any of the components in the chain between $\Us$ and $\Zs$ changes. This can be for instance the noise spectrum, the optical \ac{SNR}, but also the modulation format or the \ac{DSP} algorithms. We provide a rigorous definition of universality later in Sec.~\ref{sec:universalityrevisit}.

Unfortunately, not much is known about the universality of practical coding schemes. It is conjectured that many practical (binary) \ac{LDPC} codes are approximately universal~\cite{Franceschini06} which has been shown to be asymptotically true under some relatively mild conditions~\cite{SasonUniversal}. Guidelines for designing \ac{LDPC} codes that show good universality properties are highlighted in~\cite{sanaei2008design}. The class of spatially coupled \ac{LDPC} codes, recently investigated for optical communications~\cite{SchmalenSCJLT} has been shown to be asymptotically universal~\cite{Kudekarxx13}. An example of a \emph{non}-universal coding scheme are the recently proposed, capacity-achieving polar codes~\cite{ArikanPolar}, which need to be redesigned for every different channel. Most of these results are for binary codes and even less is known for nonbinary codes.

Although most practical \ac{LDPC} codes are asymptotically universal, we wish to emphasize a word of caution: practical, finite-length realizations of codes may only be approximately universal. For instance,~\cite[Fig. 3]{Franceschini06} reveals that for some practical LDPC codes, the performance at a \ac{BER} of $10^{-4}$ significantly differs for different channels. This difference is expected to be even larger at very low \acp{BER} due to the different slopes of the curves. We will address this difference in detail in Sec.~\ref{sec:universalityrevisit}.

\subsection{Channel Capacity and Mutual Information}

Consider an information stable, discrete-time channel with memory~\cite{verdu1994general,Agrell2014_JLT,Liga_submission16}, which is characterized by the sequence of \ac{PDF}s $p_{\bm{\Ys}_1^N | \bm{\Xs}_1^N}(\bm{y}_1^N|\bm{x}_1^N)$, for $N=1,2,\ldots$.
The maximum rate at which reliable transmission over such a
channel is possible is defined by the \emph{channel capacity}~\cite{verdu1994general,Agrell2014_JLT,Liga_submission16}
\begin{align}\label{C}
C := \lim_{N\to\infty}\sup_{p_{\bm{\Xs}_1^N}}\frac{1}{N}I(\bm{\Xs}_1^N; \bm{\Ys}_1^N)
\end{align}
where the maximization is over $p_{\bm{\Xs}_1^N}(\cdot)$, which is the \ac{PDF} of the sequence $\bm{\Xs}_1^N = (\Xs[1], \Xs[2],\ldots, \Xs[N])$ under a given input constraint (e.g., power constraint). For a fixed $p_{\bm{\Xs}_1^N}(\cdot)$, the \acf{MI} between the input sequence $\bm{\Xs}_1^N$ and the output sequence $\bm{\Ys}_1^N$ is given by
\[
I(\bm{\Xs}_1^N; \bm{\Ys}_1^N) = \mathbb{E}_{p_{\bm{\Xs}_1^N,\bm{\Ys}_1^N}}\left\{\log_2\frac{p_{\bm{\Ys}_1^N|\bm{\Xs}_1^N}(\bm{\Ys}_1^N|\bm{\Xs}_1^N)}{p_{\bm{\Ys}_1^N}(\bm{\Ys}_1^N)}\right\}
\]
where $\mathbb{E}_{p_{{R}}}\{f({R})\} := \int_{\text{dom}({R})}p_{{R}}(r)f(r){\rm d}r$ denotes expectation with respect to a random variable ${R}$.

The capacity $C$ in \eqref{C} is the maximum information rate that can be achieved for any transmission system, requiring carefully optimized, infinitely long input sequences. Usually, in most of today's systems, the channel input sequence is heavily constrained (e.g., by the use of \ac{QAM} constellations) to simplify the transceiver design. Furthermore, often symbol sequences with \ac{IID} elements are used such that we have
\begin{align}\label{product}
p_{\bm{\Xs}_1^N}(\bm{x}_1^N) = \prod_{i=1}^NP_{\Xs}(x[i]) = \prod_{i=1}^N \lambda_{\phi^{-1}(x[i])}.
\end{align}
\ac{IID} symbol sequences are obtained if a memoryless mapper is used (as we do in this paper, see, e.g., $\Phi$ in Fig.~\ref{fig:system_model}) and if sufficiently long interleaving is applied after \ac{FEC} encoding. Under these conditions, an \acf{AIR} is given by
\begin{align}\label{I.mem}
I_{\text{mem}} = \lim_{N\to\infty} \frac{1}{N}I(\bm{\Xs}_1^N; \bm{\Ys}_1^N) \leq C
\end{align}
which is a lower bound to the capacity $C$ due to the constraints imposed on the transmitted sequences. In the remainder of this paper, we limit ourselves to \ac{IID} channel input sequences generated via \eqref{product}. 

The numerical evaluation of the \ac{MI} in \eqref{I.mem} is in general not practical. The reasons are as follows: First, numerically evaluating $I(\bm{\Xs}_1^N;\bm{\Ys}_1^N)$ is hard, even for for relatively short input and channel output sequences (small memory lengths $N$). Second, most of today's transceivers do not exploit memory but instead use long interleavers to remove all effects of memory to keep decoding simple with symbol-by-symbol detection. Hence, it would not be fair to provide thresholds based on memory, which give a performance that could be achieved at some point in the future, provided that all memory is adequately exploited at the transceiver. Instead, we neglect all memory effects and obtain thresholds that indicate a performance \emph{achievable} with today's systems.
  
 Therefore in this paper, we focus on symbol-by-symbol detection (see Fig.~\ref{fig:system_model}). Under these constraints, we can further lower bound the \ac{MI} in \eqref{I.mem} (see~\cite[Sec. III-F]{essiambre2010capacity} for an in-depth proof) by employing a memoryless channel transition \ac{PDF} $p_{\Ys|\Xs}(\cdot|\cdot)$ that is obtained by averaging the true channel \ac{PDF}. This approach gives
\begin{align}
I(\Xs; \Ys) &= \mathbb{E}_{p_{\Xs,\Ys}}\left\{\log_2\frac{p({\Ys}|{\Xs})}{p({\Ys})}\right\} \leq I_{\text{mem}} \leq C\label{eq:midef}
\end{align}
or equivalently
\begin{multline}
I(\Xs;\Ys) = \\
\sum_{i=1}^{M}\lambda_{i}\int\limits_{{y}\in\mathbb{R}^D}p_{\Ys|\Xs}(y|s_i)\log_2\left(\frac{p_{\Ys|\Xs}(y|s_i)}{\sum_{j=1}^Mp_{\Ys|\Xs}(y|s_j)\lambda_{j}}\right){\rm d}y.\label{eq:sd_MI}
\end{multline}
Note that $I(\Xs;\Ys)$ is an \ac{AIR} for systems employing optimum decoding, i.e., when the \ac{LLR} computation uses $q_{\Ys|\Xs}(y|x) = p_{\Ys|\Xs}(y|x)$, and if sufficiently long symbol-wise interleaving is applied (within the equivalent ``optical channel'') and sufficiently long capacity-achieving \ac{FEC} codes are used.

\section{Thresholds for Nonbinary FEC}\label{sec:thresholds}

Based on the discussion in Sec.~\ref{sec:universality}, here we propose to use the \ac{MI} as performance thresholds for \ac{NB}-\ac{FEC}. After a discussion on how to compute these thresholds, we describe some other commonly used thresholds.

\subsection{Thresholds Based on Mutual Information}

In order to estimate the performance of \ac{NB}-\ac{FEC}, motivated by the universality argument in Sec.~\ref{sec:universality}, we would like to use the \ac{MI} $I(\Us;\Zs)$ as performance threshold. $I(\Us;\Zs)$ is the \ac{MI} between the FEC encoder output $\Us$ and FEC decoder input $\Zs$ (see Fig.~\ref{fig:system_model}) and characterizes the nonbinary channel. Unfortunately, the \ac{MI} $I(\Us;\Zs)$ is not easy to compute immediately, which is why we define a threshold that is directly related to the input $X$ and output $Y$ of the optical transmission experiment, to which we usually have access. This also allows us to avoid including soft symbol demodulation in the transmission experiment.

In the previous section, we have seen that $I_{\text{mem}}$ is a maximum \ac{AIR} if all memory effects are taken into account and is an upper bound on $I(X;Y)$, which is an \ac{AIR} under optimum decoding with an averaged channel \ac{PDF}. As a consequence of the data processing inequality, we have
\[
I_{\text{mem}} \geq I(X;Y) \stackrel{(a)}{\geq} I(U;Z)
\]
where we have equality in $(a)$ only in some special cases described below. Due to this inequality, we cannot always directly use $I(X;Y)$ as a proxy for estimating $I(U;Z)$. We resort to the theory of mismatched decoding~\cite{gantimismatched}\cite{merhav1994information} and propose to use $\underline{I}(\Xs;\Ys)$ as estimate of $I(U;Z)$, where
\begin{multline}
\underline{I}(\Xs;\Ys) := \\
\sup_{\nu \geq 0}\mathbb{E}_{p_{\Xs,\Ys}}\left\{\log_2\left(\frac{[q_{\Ys|\Xs}(\Ys|\Xs)]^\nu}{\sum_{j=1}^M\lambda_{j} [q_{\Ys|\Xs}(\Ys|s_j)]^\nu}\right)\right\}\,.\label{eq:modgmi}
\end{multline}
We have $I(X;Y) \geq \underline{I}(\Xs;\Ys) \geq I(U;Z)$, where the second inequality is due to~\cite{gantimismatched,merhav1994information}. However, we found in numerical simulations and in transmission experiments that, in the context of optical communications, $\underline{I}(\Xs;\Ys)\approx I(U;Z)$. Hence, we can use $\underline{I}(\Xs;\Ys)$ as an accurate estimate of $I(\Us;\Zs)$ and of the \ac{NB}-\ac{FEC} performance.

Even~\eqref{eq:modgmi} is demanding to evaluate in general, as the expectation is taken over $P_{\Ys,\Xs}(y,x) = p_{\Ys|\Xs}(y|x)\lambda_{\phi^{-1}(x)}$, which is often not known. However, we can replace the expectation in~\eqref{eq:modgmi} by the empirical average, as done for instance in~\cite[Sec.~III]{buchali_jlt15}. We denote this empirical approximation of $\underline{I}(\Xs;\Ys)$ by $I_{\text{NB}}$, which can be computed from an optical transmission experiment with a measurement database of $N_m$ measured values $x[\kappa]\in\mathcal{S}$ and their corresponding received $y[\kappa]$ by
\begin{multline}
I_{\text{NB}} := \frac{1}{N_m}\sup_{\nu \geq 0}\sum_{\kappa=1}^{N_m}\log_2\left(\frac{[q_{\Ys|\Xs}(y[\kappa]|x[\kappa])]^\nu}{\sum_{j=1}^M \lambda_{j}[q_{\Ys|\Xs}(y[\kappa]|s_j)]^\nu }\right)\,,\label{eq:gmiempirical}
\end{multline}
where $q_{\Ys|\Xs}(y|x)$ is the same \ac{PDF} used for computing the \acp{LLR} in~\eqref{eq:llrcomp}, e.g., the $D=2$-dimensional Gaussian \ac{PDF}. The variance of this distribution can for instance be estimated from the measurement database (or a subset thereof), see, e.g.,~\cite[Sec.~III]{buchali_jlt15}. Later, in Example~\ref{ex:var}, we show how we can jointly estimate the \ac{MI} and the noise variance, avoiding an extra variance estimator.
As the optimization in~\eqref{eq:modgmi} and~\eqref{eq:gmiempirical} is over a strictly unimodal ($\cap$-convex) function in $\nu$~\cite[Thm. 4.22]{Alvarado15_Book}, the maximization can be efficiently carried out using, e.g., the Golden section search~\cite{GoldenMethod}.

\subsection{Detailed Description of the Proposed Threshold $\underline{I}(\Xs;\Ys)$}

In the following, we describe in detail the steps that lead us to the performance metric in~\eqref{eq:gmiempirical} starting from $I(\Us;\Zs)$. The remainder of this section may be skipped in a first reading.
The input $Z$ to the FEC decoder consists of vectors of $M-1$ dimensional \acp{LLR}, whose distributions are hard to estimate, especially if $M$ becomes large. Therefore, we would like to relate $I(\Us;\Zs)$ to $\Xs$ and $\Ys$, to which we have immediately access as input and output parameters of the optical transmission experiment.  Using the data processing inequality~\cite{CoverThomas}, we can bound $I(\Us;\Zs)$ as follows 
\begin{align*}
I(\Us;\Zs) &\stackrel{(a)}{\leq} I(\Xs;\Zs) \stackrel{(b)}{\leq} I(\Xs;\Ys)
\end{align*}
where we have equality in $(a)$, if the mapper $\Phi$ is a one-to-one function (this is not the case for many-to-one mappings, used in, e.g., some probabilistic shaping implementations~\cite{yankov_constellation_2014}). In this paper, we only consider one-to-one mapping functions and thus have $I(\Us;\Zs) = I(\Xs;\Zs)$. We have equality in $(b)$ if and only if $Z$ constitutes a sufficient statistic for $\Xs$ given $\Ys$~\cite{richardsonmodern}, i.e., if $\Xs$ is independent of $\Ys$ given $\Zs$. 

While equality in $(a)$ is obtained in most communication systems, we do not necessarily have equality in $(b)$, especially if we employ a mismatched decoder, i.e., when the \ac{PDF} $q_{\Ys|\Xs}(y|x)$ assumed in the decoder does not exactly correspond to the average channel \ac{PDF} $p_{\Ys|\Xs}(y|x)$. Therefore, we cannot directly use $I(\Xs;\Ys)$ but need to find a more accurate estimate of $I(\Us;\Zs)$ based on $\Xs$ and $\Ys$.

Unfortunately, in general, $p_{\Ys|\Xs}$ is not known and must be estimated from the experiment. As the noise in uncompensated coherent optical fiber communication tends to be Gaussian~\cite{poggiolini2012gn}, a good choice is to approximate $p_{\Ys|\Xs}(y|x)$ by a Gaussian \ac{PDF}, with different levels of refinement~\cite{Eriksson_MI_JLT}. In most cases, circularly symmetric Gaussian \acp{PDF} are enough, which is what we have used in~\eqref{q.awgn}. To get a more accurate estimate of the conditional channel \ac{PDF}, we can also use a \ac{KDE}~\cite{silverman1986density} to approximate the~\ac{PDF}.

As estimating the \ac{PDF} $p_{\Ys|\Xs}(y|x)$ is not always straightforward and because we may use a mismatched decoder with $I(\Us;\Zs)\leq I(\Xs;\Ys)$, we propose to use $\underline{I}(\Xs;\Ys) \leq I(\Xs;\Ys)$ given in~\eqref{eq:modgmi} as performance predictor, which originates from~\cite{gantimismatched}, and which we found to accurately predict  $I(\Us;\Zs)$ and hence the \ac{NB}-\ac{FEC} performance.

In the optical communications literature, the  \emph{auxiliary channel lower bound}\cite{arnold2006simulation}, is frequently used to estimate the MI~\cite{Eriksson_MI_JLT}\cite[Sec. III]{buchali_jlt15}\cite[Sec.~2]{fehenberger2015achievable}\cite{secondini2013achievable} and which is given by
\begin{align}
\doubleunderline{I}(\Xs;\Ys) &:= \mathbb{E}_{p_{\Xs,\Ys}}\left\{\log_2\left(\frac{{q}_{\Ys|\Xs}(\Ys|\Xs)}{\sum_{j=1}^M{q}_{\Ys|\Xs}(\Ys|s_j)\lambda_{j}}\right)\right\}\label{eq:aclb} \\
&\leq I(\Xs;\Ys)\nonumber.
\end{align}
The expectation in~\eqref{eq:aclb} is taken over the actual (averaged) joint channel \ac{PDF} $p_{\Xs,\Ys}(\cdot,\cdot)$ and ${q}_{\Ys|\Xs}(\cdot|\cdot)$ is an \emph{auxiliary} \ac{PDF}. If ${q}_{\Ys|\Xs}(y|x) = p_{\Ys|\Xs}(y|x)$, we have $\doubleunderline{I}(\Xs;\Ys) = I(\Xs;\Ys)$. Note that~\eqref{eq:aclb} is just a special case of~\eqref{eq:modgmi}  with $\nu=1$ and hence
\[
\doubleunderline{I}(\Xs;\Ys) \leq \underline{I}(\Xs;\Ys) \leq I(\Xs;\Ys)\,,
\]
where the first inequality is obvious as $\doubleunderline{I}(\Xs;\Ys)$ is recovered for $\nu=1$ in~\eqref{eq:modgmi} and the second inequality is shown in~\cite{gantimismatched}.

It is often claimed in the above-mentioned references that one should use the same ${q}_{\Ys|\Xs}(y|x)$ as we use in the decoder (e.g., to compute the \acp{LLR} in~\eqref{eq:llrcomp}) to estimate the \ac{MI} 
via~\eqref{eq:aclb}. However, we found in numerical experiments that $\doubleunderline{I}(\Xs;\Ys)$ can significantly underestimate $I(\Us;\Zs)$ in many practical applications. We illustrate this discrepancy by means of an example.
\begin{example}
Consider the following toy example for $D=1$ where $p_{\Ys|\Xs}(y|x) = \mathcal{N}(x,\sigma_n^2)$, 
i.e., is Gaussian distributed with variance $\sigma_n^2$ and mean $x$ and where $q_{\Ys|\Xs}(y|x) = \mathcal{N}(x,K)$, i.e., the receiver assumes a Gaussian distribution with different variance $K\neq\sigma_n^2$. In this case, we can show that $I(\Xs;\Ys) = I(\Us;\Zs)$, as we can represent $p_{\Ys|\Xs}(y|x) = a(x,z)b(y)$~\cite[Sec. 1.10]{richardsonmodern}\cite[Lem. 4.7]{richardsonmodern}. The random variable $\Zs$ is an $M-1$ dimensional vector with entries $Z_i$ and realizations $z_i$. Let $z_i = \log\left(\frac{q_{\Ys|\Xs}(y|s_i)}{q_{\Ys|\Xs}(y|s_1)}\right)$. We can thus write, for $i\in\{1,\ldots M\}$,
\begin{multline*}
q_{\Ys|\Xs}(y|x) = \\
q_{\Ys|\Xs}(y|s_1)\exp\left((1-\mathbbm{1}_{\{x=s_1\}})\sum_{i=2}^Mz_{i-1}\mathbbm{1}_{\{x=s_i\}}\right)
\end{multline*}
where $\mathbbm{1}_{\{\cdot\}}$ is the binary indicator function. Relating $p_{\Ys|\Xs}(y|x)$ to $q_{\Ys|\Xs}(y|x)$ yields
\begin{multline*}
(\sqrt{2\pi\sigma_n^2})^{\frac{\sigma_n^2}{K}-1}\sqrt{\frac{\sigma_n^2}{K}}\left[p_{\Ys|\Xs}(y|x)\right]^{\frac{\sigma_n^2}{K}} = \\
q_{\Ys|\Xs}(y|s_1)\exp\left((1-\mathbbm{1}_{\{x=s_1\}})\sum_{i=2}^Mz_{i-1}\mathbbm{1}_{\{x=s_i\}}\right)
\end{multline*}
which allows us to write
\begin{multline*}
p_{\Ys|\Xs}(y|x) = \underbrace{\exp\left(\frac{K}{\sigma_n^2}(1-\mathbbm{1}_{\{x=s_1\}})\sum_{i=2}^Mz_{i-1}\mathbbm{1}_{\{x=s_i\}}\right)}_{=:a(x,z)}\times \\
\times \underbrace{\left(\frac{K}{\sigma_n^2}\right)^{\frac{K}{2\sigma_n^2}}\left(\sqrt{2\pi\sigma_n^2}\right)^{\frac{K}{\sigma_n^2}-1}\left[q_{\Ys|\Xs}(y|s_1)\right]^{\frac{K}{\sigma_n^2}}}_{=:b(y)}
\end{multline*}
and hence we have $I(\Xs;\Ys) = I(\Xs;\Zs)$. However, if we evaluate $\doubleunderline{I}(\Xs;\Ys)$ from~\eqref{eq:aclb} for $K\neq \sigma_n^2$, we inevitably have $\doubleunderline{I}(\Xs;\Ys) < I(\Xs;\Ys)$.

If we employ for example \ac{LDPC} codes with the widely used min-sum decoder or the less known linear programming decoder~\cite{feldman2005}, the decoding performance does not depend on $K > 0$ used for computing the \acp{LLR} and hence, $\doubleunderline{I}(\Xs;\Ys)$ will not be an adequate performance estimate and may even largely underestimate the performance, if used as performance prediction threshold.  \hfill $\bigtriangleup$
\end{example}

We therefore propose to use the generalization~\eqref{eq:modgmi} of~\eqref{eq:aclb}, which  we found to accurately predict  $I(\Us;\Zs)$ and hence the \ac{NB}-\ac{FEC} performance. 
A convenient byproduct of using $\underline{I}(\Xs;\Ys)$ is the fact that it can be used to jointly estimate the \ac{MI} $I(\Us;\Zs)$ and the variance of the noise. We illustrate this application in the following example.

\begin{example}\label{ex:var}
For the case of uncompensated links, we know that the Gaussian \ac{PDF} is a good approximation of the channel \ac{PDF}~\cite{poggiolini2012gn}. However, in general, as we do not know \emph{a priori} the variance of the noise \ac{PDF}, we need to estimate it. In~\cite[Sec. III]{buchali_jlt15}, it is for instance proposed to estimate the noise variance from the measurement database. Here we propose to directly use the \ac{MI} estimate to obtain the noise variance. As the variance is unknown, we first fix $\sigma_n^2 = \frac{1}{2}$ in~\eqref{q.awgn} and then evaluate~\eqref{eq:gmiempirical} as
\begin{multline}
I_{\text{NB}}\Big\vert_{\textsf{awgn}} = \frac{1}{\ln(2)N_m}\sup_{\nu \geq 0}\sum_{\kappa=1}^{N_m}\Bigg(-\nu\left\lVert y[\kappa]-x[\kappa]\right\rVert^2 -  \\
\left. \log\left(\sum_{j=1}^M\lambda_j\exp\left(-\nu\left\lVert y[\kappa]-s_j\right\rVert^2\right)\right)\right).\label{eq:ex2}
\end{multline}
After carrying out the optimization over $\nu$ (for example, as highlighted above, using the Golden section search), we immediately get an estimate of the noise variance as $\hat{\sigma}_n^2 =  \frac{1}{2\hat{\nu}}$, where $\hat{\nu}$ is the $\nu$ that maximizes~\eqref{eq:ex2}.\hfill $\bigtriangleup$
\end{example}

\subsection{Other Thresholds}\label{sec:thresholdsref}

In the remainder of this paper, the accuracy of the \ac{MI} as a decoding threshold will be compared against predictions based on other performance thresholds.
If \ac{BICM}, as explained in Sec.~\ref{sec:bicm} and shown in Fig.~\ref{fig:system_model_bicm}, is used as \ac{CM} scheme, the \emph{bit-wise} \ac{GMI} is a good metric~\cite{Alvarado2015b_JLT}. The \ac{GMI} is computed as
\begin{multline*}
\text{GMI} \approx -\frac{1}{N_m}\sum_{\kappa=1}^{N_m}\log_2(\lambda_{\phi^{-1}(y[\kappa])}) - \\
\frac{1}{N_m}\sum_{i=1}^m\sum_{\kappa=1}^{N_m}\log_2\left(1+\exp\left((-1)^{c_{i}[\kappa]}\tilde{L}_i(y[\kappa])\right)\right),
\end{multline*}
where $c_i[\kappa]$ is the bit at bit position $i$ ($i\in\{1,\ldots, m\}$) mapped to symbol $x[\kappa]$ and $\tilde{L}_i(y[\kappa])$ are the \emph{bit-wise} \acp{LLR} computed according to~\eqref{eq:llr_bicm}. In the practically prevalent case where all symbols are equiprobable ($\lambda_i = 1/M$), we have
\[
\text{GMI} \approx m - \frac{1}{N_m}\sum_{i=1}^m\sum_{\kappa=1}^{N_m}\log_2\left(1+\exp\left((-1)^{c_{i}[\kappa]}\tilde{L}_i(y[\kappa])\right)\right).
\]
 We assume from now on for simplicity that all constellation symbols are equiprobable, i.e., $\lambda_i = 1/M$. Computing the \ac{GMI} in a nonbinary transmission scheme necessitates the use of an extra \emph{LLR Calculator} implementing~\eqref{eq:llr_bicm}, which is shown in the bottom branch of Fig.~\ref{fig:system_model}.

Second, we use the pre-FEC~\ac{BER} $\frac{1}{m}\sum_{i=1}^{m}P(\hat{B}_i\neq B_i)$, and the pre-FEC~\ac{SER} $P(\hat{X}\neq X)$. These quantities are schematically shown at the bottom of Fig.~\ref{fig:system_model}. We immediately see that only the \ac{MI} is directly connected to the {\acs{NB}-\ac{FEC}} decoder, and thus is the most natural threshold choice. In particular, the transmitter in Fig.~\ref{fig:system_model} uses a GF($2^{m}$)-to-bit mapper followed by a bit-to-symbol mapper $\Phi(\boldsymbol{b})=x$, which maps the vector of bits $\boldsymbol{b}=(b_1,b_2,\ldots, b_m)$ to a constellation symbol $x\in\mathcal{S}$. These blocks are included only so that the \ac{GMI} and pre-FEC~\ac{BER} can be defined (and calculated) but have no operational significance for the \acs{NB}-\ac{CM} system under consideration, as $U$ can be directly mapped to $X$. The bit labeling used in the mapper $\Phi$ affects both the \ac{GMI} and pre-FEC~\ac{BER}, but has no impact on the actual performance of the system.
At the receiver side, additionally logarithmic likelihood ratios (LLRs) are calculated ($\boldsymbol{\tilde{L}}$), and a hard-decision on the symbols is made ($\hat{X}$), which  leads to a hard-decision on the bits ($\hat{\boldsymbol{B}}$).

\subsection{Performance Prediction for \ac{BICM}-ID and Multi-Level Coding}
An alternative to \ac{BICM} is to use \ac{BICM}  with iterative demapping (BICM-ID), a concept introduced in \cite{Li97,Brink98}. The idea is to use iterative demapping to compensate for the information loss from non-ideal \ac{BMD}. \ac{BICM}-ID for optical communications has been studied for instance in \cite{Djordjevic2007_JLT,Batshon2009_JLT, Bulow14}, \cite[Sec.~3]{Bulow2011b}, \cite[Sec.~3]{Bulow2011}, \cite[Sec.~4]{Schmalen14}. In BICM-ID, iterations between the decoder and demapper are added to a possibly already iterative FEC decoder. To keep the number of iterations low, however, one can trade FEC decoder iterations for demapper iterations. The design of \ac{BICM}-ID is more complex than \ac{BICM}, however, \ac{BICM}-ID is expected to perform very close to a \ac{ML} sequence detector, and thus, to outperform \ac{BICM}. 

The \ac{MI}, as described in this section, is supposedly also a good performance estimator for \ac{BICM}-ID systems. However, while \ac{BICM} schemes with commonly used \ac{FEC} implementation behave fairly universal (see also Sec.~\ref{sec:universality} and~\ref{sec:universalityrevisit}), we found that this is not the case with \ac{BICM}-ID. Even small changes in the channel or the modulation format can cause severe differences in the performance of \ac{BICM}-ID schemes. For example, in~\cite{SchmalenOFC15,SchmalenAdvances}, we have shown that in systems with iterative differential detection for optical systems affected by phase slips, even a change of the phase slip probability can lead to significant performance differences. In \ac{BICM}-ID, generally, the \ac{FEC} code has to be optimized for every modification of channel and modulation format, i.e., the universality is not guaranteed.  Therefore, we suggest to always explicitly carry out decoding in \ac{BICM}-ID systems, as shown in, e.g., in~\cite{SchmalenOFC12} or to use \ac{MI} thresholds that have been obtained with a simulation reflecting exactly the setup of the experiment.

Recently, we have shown that the use of spatially coupled (SC) \ac{LDPC} codes~\cite{SchmalenSCJLT} can lead to a more universal behavior when used as \ac{FEC} schemes in \ac{BICM}-ID~\cite{SchmalenSCC13,SchmalenOFC15,SchmalenAdvances}. These results are however still preliminary and mostly based on asymptotic arguments. First simulations successfully demonstrated the improved universality of SC \ac{LDPC} codes.

The same argument also applied to \ac{MLC} with \ac{MSD}~\cite{wachsmann1999multilevel}. This scheme is capacity-achieving and hence, the \ac{MI} is a good performance estimator. However, \ac{MLC} with \ac{MSD} is intrinsically nonuniversal and the selection of code rates has to be adapted for every change of channel, modulation format, and bit mapping~\cite{wachsmann1999multilevel}, which is why we also recommend either to carry out decoding or to use an \ac{MI}-based threshold which has been obtained from simulations of a setup identical to the one used in the transmission experiment.

\section{Experimental Verification}\label{sec:experiment}

\begin{figure}[t!]
\centering
\includegraphics{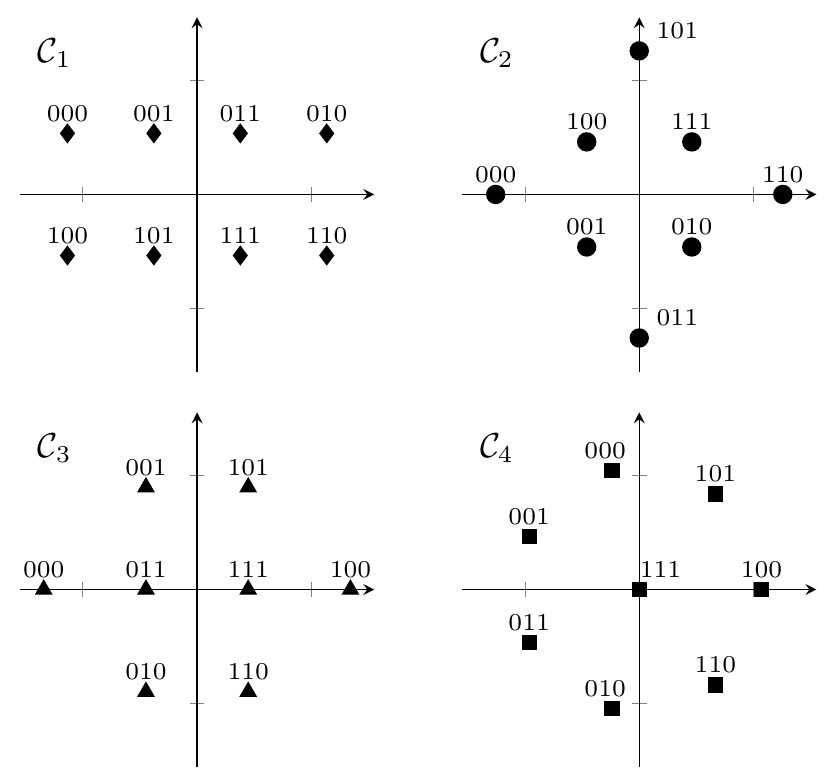}
\vspace*{-2ex}
\caption{Four different 8-QAM constellations used in the numerical results taken from~\cite{RiosMullerQAM8}. The numbers adjacent to the constellation points give the \ac{GMI}-maximizing bit labeling. The markers used for the constellation points will be subsequently used to distinguish the constellations.}\label{fig:constellations}
\end{figure}

To experimentally verify the proposed method, we consider the four 8-QAM constellations shown in Fig.~\ref{fig:constellations}, where the bit-mapping that maximizes the \ac{GMI} is also shown~\cite{RiosMullerQAM8}\cite{RiosMullerECOC14}. For illustration purposes, we use five quasi-cyclic \acs{NB}-\ac{LDPC} codes with rates $R\in\{0.7,0.75,0.8,0.85,0.9\}$ (FEC overheads of $\approx 43,33,25,18,11$\%) defined over GF($2^3$) with regular variable node degree of $d_v=3$ and regular check node degrees $d_c\in\{10,12,15,20,30\}$ of girth 8 ($R<0.9$) or girth 6 ($R=0.9$), respectively. Each code has length of around $5500$, i.e., always $5500$ 8-QAM symbols are mapped to one \ac{LDPC} codeword. The parameters of the codes are summarized in Tab.~\ref{tab:MIthresholds}. As the Galois field over which these codes are defined is rather small, the decoding complexity is relatively small as well. Decoding takes place using $15$ iterations with a row-layered belief propagation decoder. These codes are conjectured to be approximately universal, i.e., their performance is expected to be independent of the actual channel (see also Sec.~\ref{sec:universality}). 

Note that in the following we often use only a subset of constellations and code rates to keep the visualization of results simple and as we reuse previously recorded measurements. Note that the main purpose of this paper is to show that we can reuse previously recorded experimental data and evaluate the performance of \ac{NB}-\ac{FEC} for these experiments which is why we avoid redoing experiments.

\begin{figure*}[tbh!]
\centering
\includegraphics{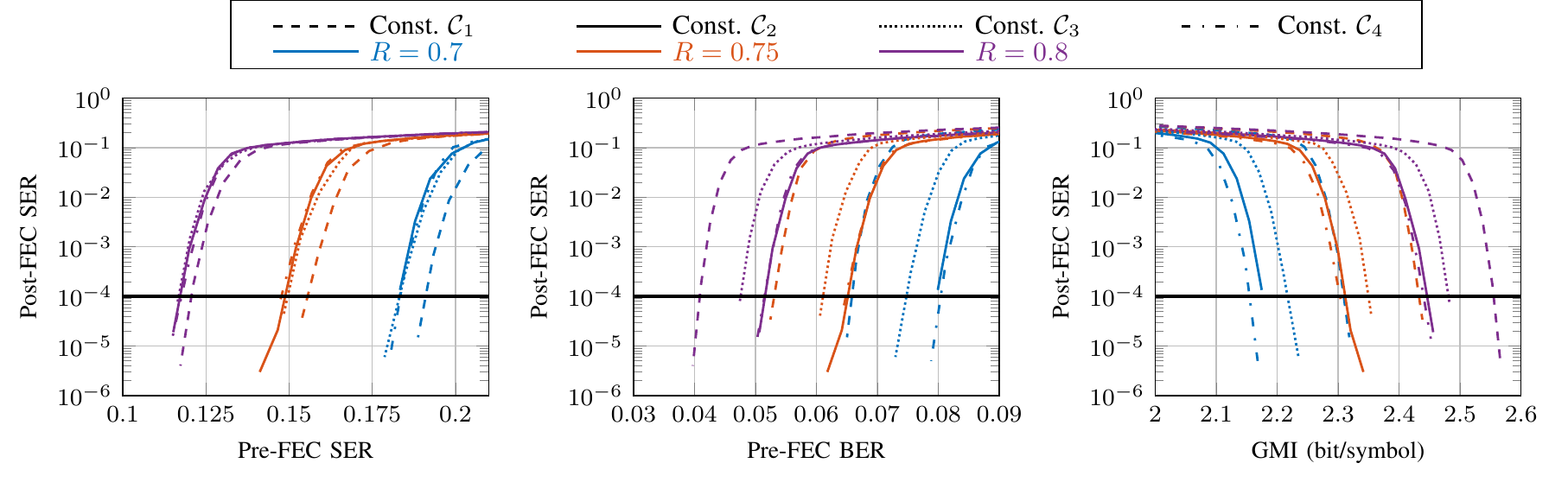}
\vspace*{-5ex}
\caption{Post-FEC SER as a function of three different performance metrics (pre-FEC SER, pre-FEC BER and GMI) for three NB-LDPC codes.}
\label{fig:comparison}
\end{figure*}

\begin{table}[bt!]
\centering
\caption{Code parameters and MI thresholds $T_R$ for different code rates $R$}\label{tab:MIthresholds}
\begin{tabular}{@{}c|ccccc@{}}
\hline

\hline
Rate $R$ & $0.7$ & $0.75$ & $0.8$ & $0.85$ & $0.9$\\
\hline
\hline
Var. degree $d_v$ & 3 & 3 & 3 & 3 & 3\\
Check degree $d_c$ & 10 & 12 & 15 & 20 & 30\\
\hline

\hline
MI threshold $T_R$ & 2.31 & 2.43 & 2.55 & 2.67 & 2.79 \\ 
\hline
normalized MI & \multirow{2}{*}{0.77} & \multirow{2}{*}{0.81} & \multirow{2}{*}{0.85} & \multirow{2}{*}{0.89} & \multirow{2}{*}{0.93} \\
threshold ${T_R}/{m}$ & & & & &
\end{tabular}
\end{table}

\subsection{AWGN Simulation Results}

\begin{figure}[t!]
\centering
\includegraphics{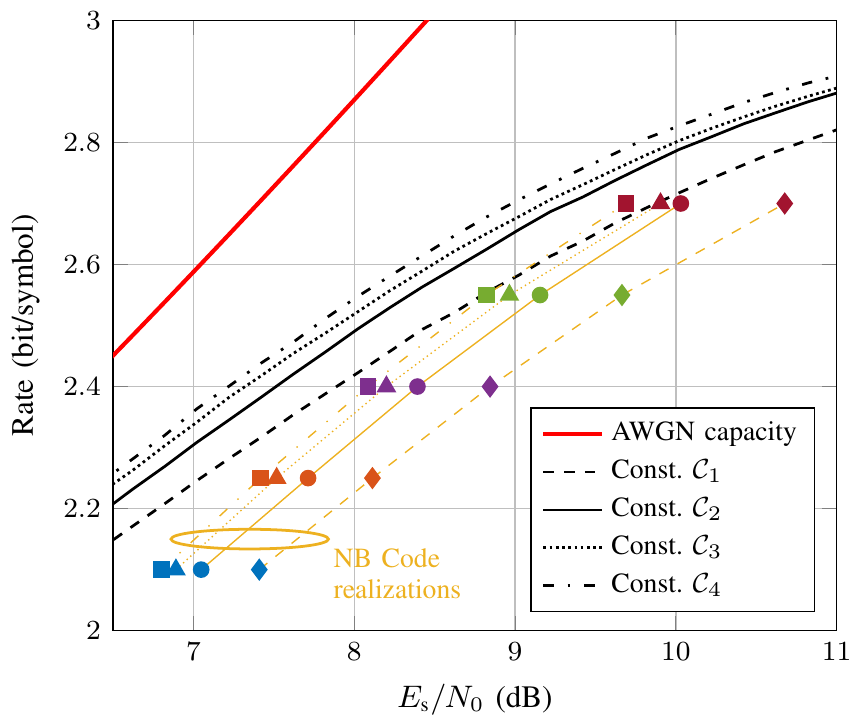}
\vspace*{-5ex}
\caption{MI (lines) and throughput (lines with markers) for the four 8-QAM constellations in Fig.~\ref{fig:constellations} and the five NB-LDPC codes in Tab.~\ref{tab:MIthresholds}. The AWGN capacity is also shown for comparison (thick red line).}\label{MI.8QAM}
\vspace*{-2ex}
\end{figure}

The performance of the five NB-LDPC codes was first tested in an AWGN channel. To this end, we first calculated the MI for the four constellations in Fig.~\ref{fig:constellations}. These results are shown as a function of  the average symbol energy-to-noise ratio $E_{\text{s}}/N_0$ in Fig.~\ref{MI.8QAM} and show a clear superiority of the constellation $\mathcal{C}_4$ in terms of MI.

In Fig.~\ref{MI.8QAM}, we also show the required $E_{\text{s}}/N_0$ for the different \ac{NB}-\ac{LDPC} codes to achieve a post-FEC \ac{SER} of $10^{-4}$ and plot that together with the corresponding net rate, given by the number of bits per constellation symbol. The obtained results show that the NB-LDPC codes follow the MI predictions quite well, although we do observe an increasing rate loss as the code rate decreases. We attribute this loss to the nonideal code design based on the fact that we only use regular codes. Optimized irregular \ac{NB}-\ac{LDPC} codes~\cite{GellerNonbinary} would be necessary for constructing better \ac{NB}-\ac{LDPC} codes at low rates.

\begin{figure}[t!]
\centering
\includegraphics{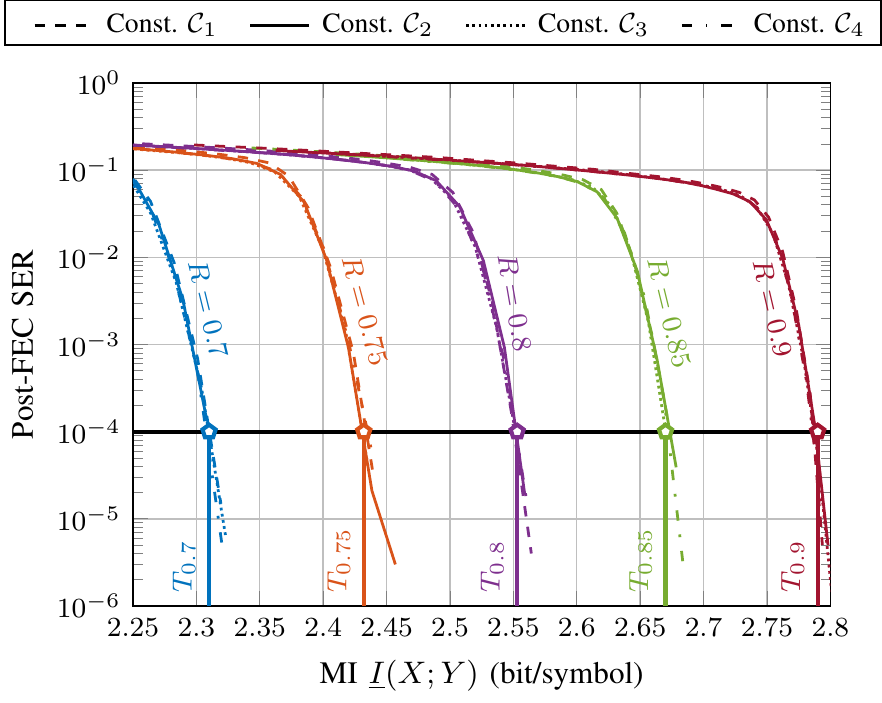}
\vspace*{-4ex}
\caption{Use of MI as performance metric for NB-LDPC codes.}
\label{fig:comparisonMI}
\vspace*{-2ex}
\end{figure}

In Fig.~\ref{fig:comparison}, we show the post-FEC~\ac{SER} as a function of the three performance metrics described in Sec.~\ref{sec:thresholdsref} for code rates $R\in\{0.7,0.75,0.8\}$. Changing the constellation for a given code can be interpreted as changing the nonbinary channel in Fig.~\ref{fig:system_model}. Additionally, in Fig.~\ref{fig:comparisonMI}, we show the proposed nonbinary MI estimate~$\underline{I}(X;Y)$ as performance metric for all four constellations and all five code rates. The results in Figs.~\ref{fig:comparison} and~\ref{fig:comparisonMI} clearly show that only the \ac{MI} can be used as a reliable threshold. In particular, for a post-FEC SER of $10^{-4}$ (horizontal lines in Figs.~\ref{fig:comparison} and~\ref{fig:comparisonMI}), the obtained MI thresholds are summarized in the third row of Tab.~\ref{tab:MIthresholds}. 

Instead of the \ac{MI}, Fig.~\ref{fig:comparison} suggests that the pre-FEC SER could also potentially serve as a performance indicator, although not as reliable as the MI. 
With the exception of constellation $\mathcal{C}_1$, the pre-FEC SER (which depends on the distance spectrum, i.e., the distances between constellation points) could be an indicator as well. Furthermore, for high rate codes, the pre-FEC SER becomes a better indicator. This is in line with the findings of~\cite{Alvarado2015b_JLT}, where it was shown that the \ac{GMI} is the proper performance indicator for systems with \ac{BICM} but for high rate codes, the pre-FEC BER can still be used with a reliability that may be good enough for some applications.

\subsection{Back-to-Back Transmission of 8-QAM Formats}

To validate the \ac{AWGN} results in Fig.~\ref{fig:comparison}, we now consider a dual-polarization $41.6$~Gbaud system. The three 8-QAM constellations of Fig.~\ref{fig:comparison} were generated and tested using a high-speed DAC in a back-to-back configuration. A root-raised cosine pulse shaping (roll-off factor $0.1$) signal was generated as described in~\cite{RiosMullerQAM8} and two code rates ($R=0.7$ and $R=0.8$) were considered, giving net data rates of approximately $174$ and $200$~Gbit/s.

\begin{figure}[t!]
\centering
\includegraphics{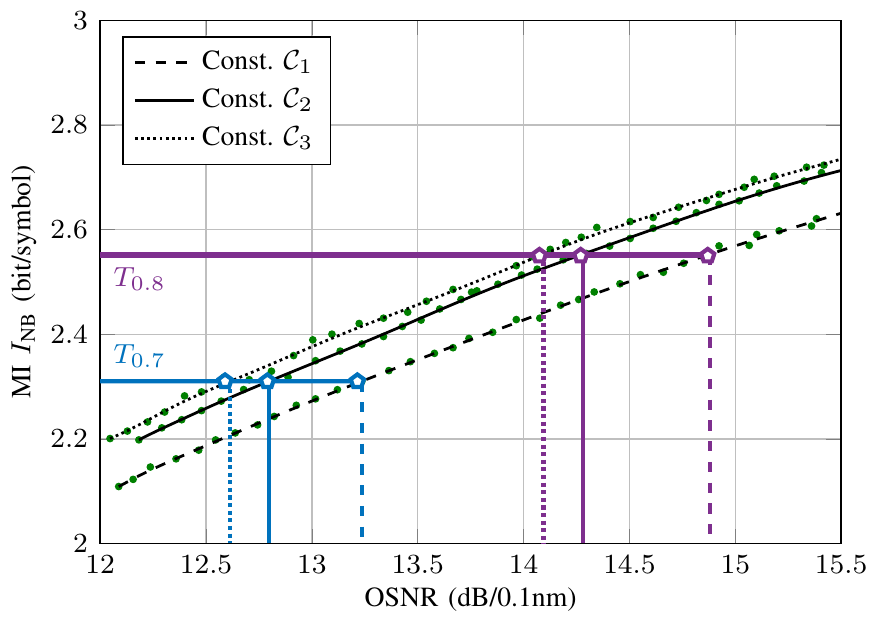}
\vspace*{-4ex}
\caption{Empirically obtained (\textcolor{black!50!green}{green markers}) and interpolated (lines) MI curves}
\label{fig:experiment_fig1}
\end{figure}

\begin{figure}[t!]
\centering
\includegraphics{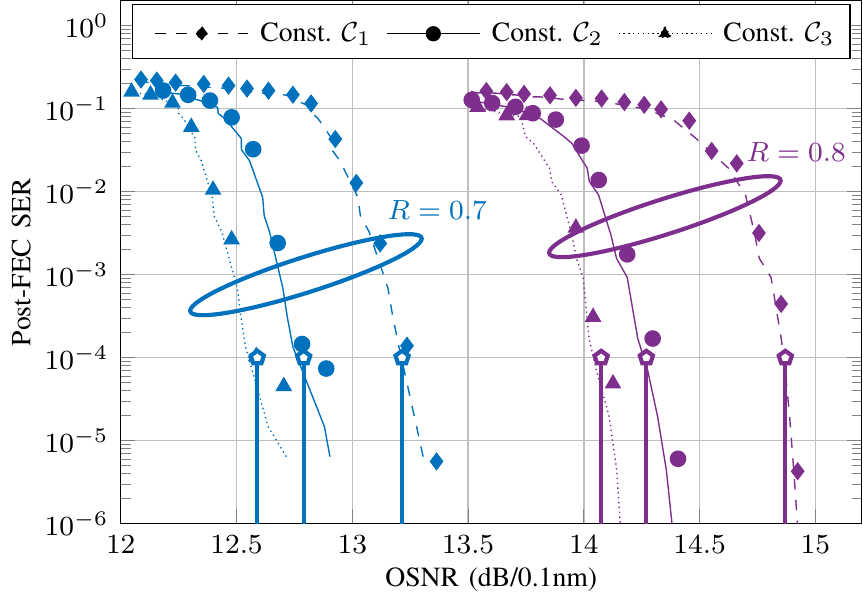}
\vspace*{-2ex}
\caption{Results after actual decoding with an \acs{NB}-\ac{LDPC} decoder with solid markers representing actual results after \ac{FEC} decoding and lines representing interpolated post-\ac{FEC} \ac{SER} estimates taken from the estimated \ac{MI}.}
\label{fig:experiment_fig3}
\end{figure}

The empirical MI estimate $I_{\text{NB}}$ as a function of the OSNR for the three constellations $\mathcal{C}_1$, $\mathcal{C}_2$ and $\mathcal{C}_3$ is shown in Fig.~\ref{fig:experiment_fig1}, where the constellation $\mathcal{C}_{3}$ shows a clear superiority in terms of \ac{MI}. In this figure, we also show the \ac{MI} thresholds $T_{0.7}=2.31$ and $T_{0.8}=2.55$ from Tab.~\ref{tab:MIthresholds}. These \ac{MI} thresholds are then used to determine equivalent OSNR thresholds for all three modulation formats (see vertical lines in Fig.~\ref{fig:experiment_fig1}). The measured data was then used to perform \acs{NB}-\ac{LDPC} decoding using a combination of the methods presented in~\cite{schmalen_generic_2012} (scramblers) and~\cite{stojanovic_reusing_2013} (interleavers). The obtained results are shown in Fig.~\ref{fig:experiment_fig3} with solid markers. Additionally, 
from the estimated MI values, we interpolated the estimated post-FEC SER values using the \ac{AWGN} simulations of Fig.~\ref{fig:comparisonMI}, which are given by thin dashed (constellation $\mathcal{C}_1$), solid (constellation $\mathcal{C}_2$), and dotted (constellation $\mathcal{C}_4$) lines. We observe a very good agreement between the predicted post-FEC SER and actual post-FEC SER values and thus a good match between the \ac{MI} thresholds obtained for the \ac{AWGN} channel and the actual performance of the codes in the experiment. 
 
\begin{figure*}[tbh!]
\centering
\includegraphics[width=1.3\columnwidth]{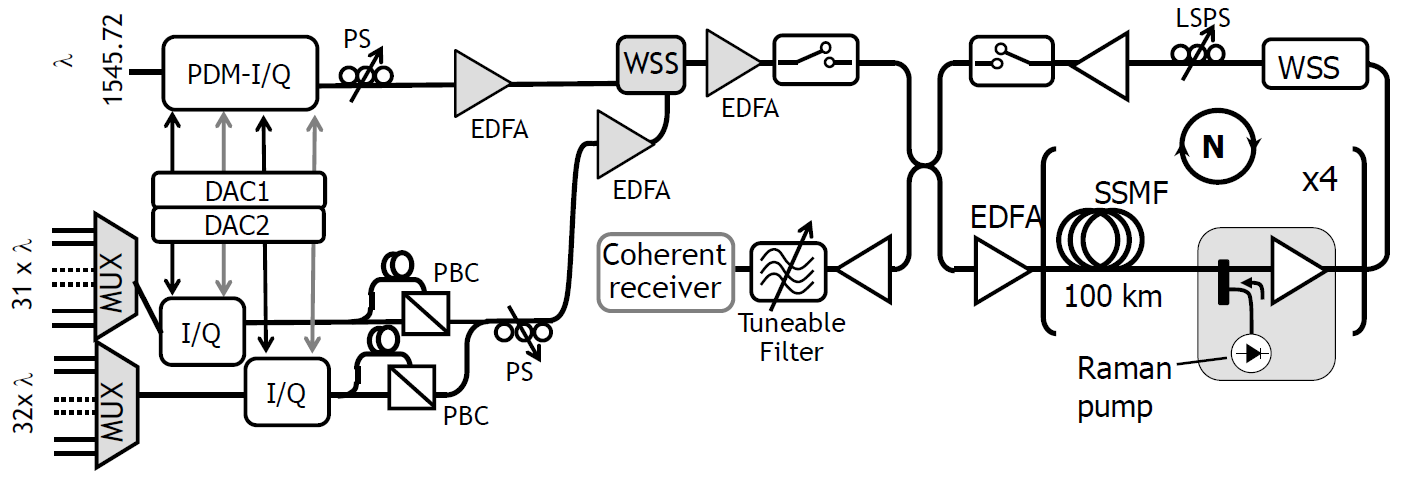}
\caption{WDM experimental setup with one channel under test, 63 WDM load channels, a recirculating loop consisting of four 100\,km spans of SSMF and hybrid Raman-EDFA amplication.}\label{fig:experimental_setup}
\end{figure*}

\subsection{Transmission Experiment}
In order to show that the proposed method also works for a transmission over a link, we apply the method to a transmission experiment using constellations $\mathcal{C}_2$ and $\mathcal{C}_4$ over a re-circulating loop, described in detail in~\cite{RiosMullerECOC14}. We recapitulate the experimental setup in the following. The transmission test-bed is depicted in Fig.~\ref{fig:experimental_setup} and consists of one narrow linewidth laser  under test at $1545.72$\,nm, and additionally 63 loading channels spaced by 50\,GHz.  The output of the laser under test is sent into a PDM I/Q modulator driven by a pair of \acp{DAC} operating at 65-GSamples/s. Multiple delayed-decorrelated sequences of $2^{15}$ bits were used to generate the multi-level drive signals.  
Pilot symbols and a sequence for frame synchronization are additionally inserted.

The symbol sequences are oversampled by a factor 
of $\approx 1.56$ and pulse shaped by a root-raised 
cosine function with roll-off of 0.1.
The load channels are separated into odd and 
even sets of channels and modulated independently 
with the same constellation as the channel under test
using separate I/Q modulators. Odd and even 
sets are then polarization multiplexed by 
dividing, decorrelating and recombining through a polarization beam combiner (PBC) with an approximate $10$\,ns 
delay. The test channel and the loading channels are  
passed into separate low-speed ($< 10$\,Hz) 
polarization scramblers (PS) and spectrally 
combined through a \ac{WSS}. The resulting multiplex is boosted 
through a single stage \ac{EDFA} and sent into the recirculating 
loop. The loop consists of four 100km-long 
dispersion uncompensated spans of \ac{SSMF}. Hybrid Raman-\ac{EDFA} optical repeaters compensate the fiber loss. The Raman 
pre-amplifier is designed to provide $\approx 10$\,dB on-off  gain.  Loop  synchronous  polarization  
scrambling (LSPS) is used and power 
equalization is performed thanks to a 50-GHz 
grid \ac{WSS} inserted at the end of the loop. 

At the receiver side, the channel under test is 
selected by a tunable filter and sent into a 
polarization-diversity coherent mixer feeding 
four balanced photodiodes. Their electrical 
signals are sampled at 80GS/s by a real-time 
digital oscilloscope having a 33-GHz electrical 
bandwidth. For each measurement, five different 
sets of 20\,$\mu s$ are stored. The received samples are processed off-line. The \ac{DSP} includes first 
chromatic  dispersion  compensation,  then  
polarization demultiplexing by a 
25-tap $T/2$ spaced butterfly equalizer with blind 
adaptation based on a multi-modulus algorithm. 

Frequency recovery is done using 4th and 7th
power periodogram for constellations $\mathcal{C}_2$ and $\mathcal{C}_4$, respectively. Phase recovery is done using the blind phase search (BPS) algorithm for both constellations. Equally-spaced 
test phases in the interval $[-\frac{\pi}{4}; \frac{\pi}{4})$ (constellation $\mathcal{C}_2$) or in the interval $[-\frac{\pi}{7}; \frac{\pi}{7}]$ (constellation $\mathcal{C}_4$) are used. The phase unwrapper is modified accordingly.

\begin{figure}[tb!]
\includegraphics{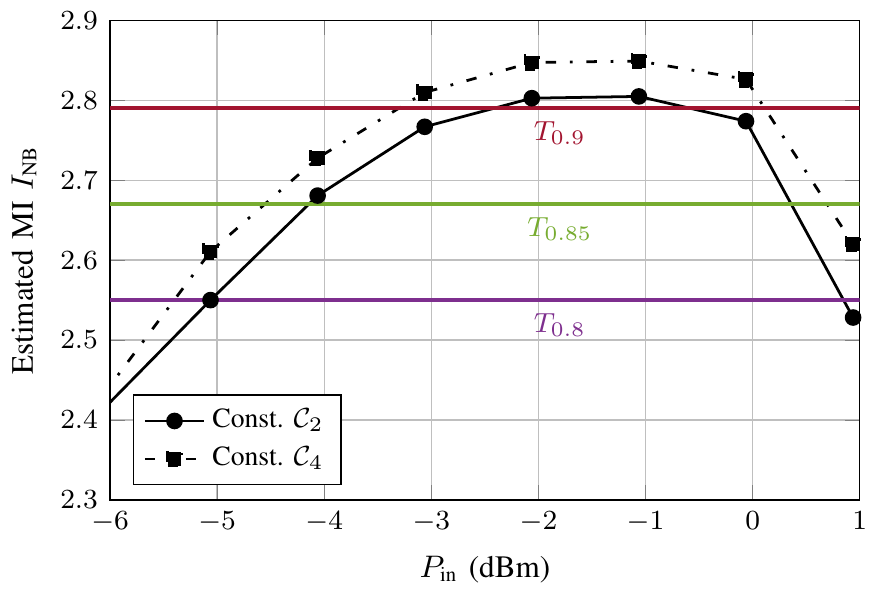}
\vspace*{-4ex}
\caption{Estimated mutual information for varying input power per channel for constellations $\mathcal{C}_2$ and $\mathcal{C}_4$ after transmission over 8 loops ($3200$\,km).}
\label{fig:experiment_threshold}
\end{figure}

We consider the transmission over 8 round trips in the recirculating loop, corresponding to a distance of $3200$\,km. Figure~\ref{fig:experiment_threshold} shows the estimated MI $I_{\text{NB}}$ as a function of the input power $P_{\text{in}}$ per \ac{WDM} channel, see also~\cite[Fig. 3-a]{RiosMullerECOC14} using the Gaussian \ac{PDF} $q_{\textsf{awgn}}\big\vert_{D=2}$ of~\eqref{q.awgn}. Using a \ac{PDF} estimate obtained with a \ac{KDE} does not lead to noteworthy differences in the \ac{MI} estimate, as predicted in~\cite{Eriksson_MI_JLT}. Additionally, we show the \ac{MI} thresholds $T_R$ for $R\in\{0.8,0.85,0.9\}$. The thresholds give us the region of launch powers at which transmission is possible.

To be precise, whenever the estimated \ac{MI} lies above the threshold $T_R$, it means that successful transmission is possible, where successful is defined in the same way as for finding the threshold, i.e., with a post-\ac{FEC} \ac{SER} below $10^{-4}$. For example, consider the red horizontal line in Fig.~\ref{fig:experiment_threshold} corresponding to $T_{0.9}$. We can see that with constellation $\mathcal{C}_2$, we are just barely above the line for $P_{\text{in}}\in\{-2\,\text{dBm},-1\,\text{dBm}\}$, which means that decoding is also only barely possible. In contrary, with constellation $\mathcal{C}_4$, we have a larger \ac{MI} margin to the threshold and therefore, reliably communication is possible over a wider range of $P_{\text{in}}$.

\begin{figure}[tb!]
\includegraphics{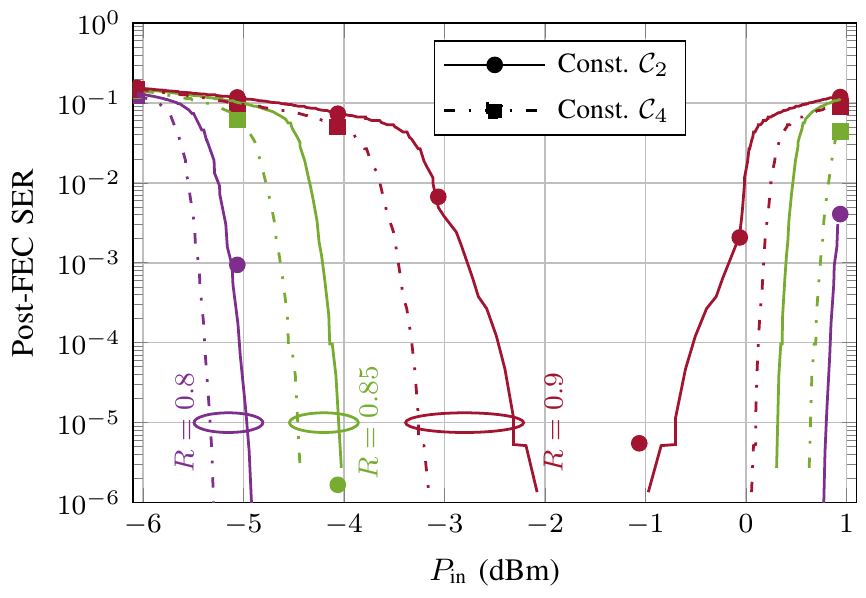}
\vspace*{-4ex}
\caption{Estimated post-FEC SER obtained by interpolation (curves) of the MI versus post-FEC SER obtained by actual decoding (markers) with LDPC codes of rate $R\in\{0.8,0.85,0.9\}$ for constellations $\mathcal{C}_2$ and $\mathcal{C}_4$ after transmission over 8 loops ($3200$\,km).}
\label{fig:experiment_interpolate}
\end{figure}

In Fig.~\ref{fig:experiment_interpolate}, we use the post-FEC \ac{SER} results of Fig.~\ref{fig:comparisonMI} to estimate the post-FEC performance of the transmission system by interpolation. The interpolated curves are given by the solid (constellation $\mathcal{C}_2$) and dash-dotted (constellation $\mathcal{C}_4$) lines. Additionally, we carried out actual decoding using the \ac{LDPC} codes introduced before. The post-FEC SER results after decoding are given by the solid markers in the figure.  We can see that the estimates from interpolation match the actual decoding performance quite well, confirming the applicability of the proposed method.

\section{Universality Revisited}\label{sec:universalityrevisit}

In the previous sections of this paper, we have seen that \ac{MI}-based thresholds can be used to accurately predict the performance of different modulation formats with the same \ac{NB}-\ac{FEC} code, for which we have computed in an offline simulation an \ac{MI}-threshold. However, we want to emphasize that caution must be taken: this approach assumes that the code is universal (see also Sec.~\ref{sec:universality}). We know from~\cite{Franceschini06} that practical codes with finite block lengths are not necessarily  universal. 

\begin{figure}[b!]
\centering
\includegraphics{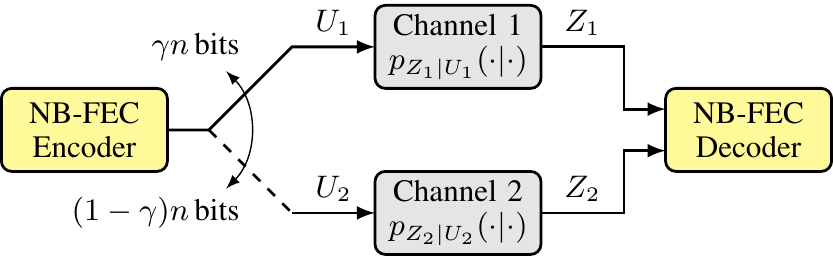}
\caption{Definition of universality of \acs{FEC} schemes according to~\cite{sanaei2008design}}
\label{fig:universalitydef}
\end{figure}

First, we give a precise definition of the concept of universality. We can define universality of \ac{FEC} schemes as in~\cite{sanaei2008design} with the help of Fig.~\ref{fig:universalitydef}. Consider an \ac{NB}-\ac{FEC} encoder that generates a codeword consisting of $n$ symbols. We consider two different communication channels with different (memoryless) channel transition \acp{PDF} $p_{Z_1|U_1}(z_1|u_1)$ and $p_{Z_2|U_2}(z_2|u_2)$, but with \emph{identical} \ac{MI} $I_C := I(U_1;Z_1) = I(U_2;Z_2)$. A fraction $\gamma n$ of the symbols is transmitted over the upper channel 1, while the remaining $(1-\gamma)n$ symbols are transmitted over the lower channel 2, where $\gamma\in[0,1]$, i.e., $\gamma$ can be any real number in the unit interval, such that $\gamma n$ is integer. We say that a code is \emph{universal} for channels 1 and 2 if the post-\ac{FEC} \ac{SER} is independent of $\gamma$. We can extend this definition to a sequence of channels and say that a code is universal if the post-\ac{FEC} \ac{SER} is independent of $\gamma$ and the channels.

\newcommand{\argmin}{\mathop{\mathrm{arg\,min}}}

In the previous examples of Sec.~\ref{sec:experiment}, we have not experienced any issue with universality, as the only changes we made in the channel were a change of the modulation format, but the underlying channel (\ac{AWGN} or optical transmission, which can be modeled accurately as \ac{AWGN}) remained fixed. In this section, we show by means of an example the impact of a more drastic change of the nonbinary channel. We now modify the channel in the \ac{AWGN} simulation by adding a hard decision to the output of the optical channel.  We assume then that the optical channel generates a hard decision output based on the Euclidean distance decision metric, i.e., the output is
\[
\hat{y}[\kappa] = s_{\hat{\imath}}\quad\text{with}\quad\hat{\imath}=\argmin_{i=1,\ldots, M}\left\lVert y[\kappa]-s_i\right\rVert.
\]
Although the outputs of the channel are \ac{NB} hard symbols, we can still carry out soft decision decoding. In soft-decision decoding, the soft symbol demodulator calculates \acp{LLR} based on the channel statistics and the received values. Assume a memoryless optical channel and let
\[
W_{j,k} := P_{\hat{\Ys}|\Xs}(s_j|s_k)
\]
denote the channel transition probability of receiving symbol $s_j$ provided that symbol $s_k$ has been sent. We can interpret this channel as a nonbinary version of the classical \ac{BSC}, often also called \ac{DMC}. We can then compute a set of \ac{NB} \acp{LLR} with
\[
L_i(\hat{y}) = \ln\left(\frac{W_{\phi^{-1}(\hat{y}),i}}{W_{\phi^{-1}(\hat{y}),1}}\right) + \ln\left(\frac{\lambda_{i}}{\lambda_{1}}\right)
\]
where $\phi(i) = s_i$ is the symbol mapping function.
We can then use these \acp{LLR} to feed a conventional soft-decision decoder. This situation may seem at a first glance counter-intuitive, as we first make a decision and then
regenerate soft-decision \acp{LLR} to use in a soft-decision \ac{NB}-\ac{FEC}. However, such a situation may arise when designing \ac{NB}-\ac{FEC} schemes for updating legacy systems that include a hard decision on symbol level which cannot be changed. The \ac{MI} for this scheme is computed as
\[
I_{\text{hd}} := I(\Xs;\hat{\Ys}) = \sum_{i=1}^M\sum_{j=1}^MW_{j,i}\lambda_{i}\log_2\left(\frac{W_{j,i}}{\sum_{k=1}^MW_{j,k}\lambda_{k}}\right).\label{I.hd}
\]
For illustration, we consider this scheme with the \ac{NB}-\ac{LDPC} codes specified in Tab.~\ref{tab:MIthresholds} and carry out a simulation over the \ac{AWGN} channel with the four 8-QAM constellations shown in Fig.~\ref{fig:constellations}.

\begin{figure}[tb!]
\centering
\includegraphics{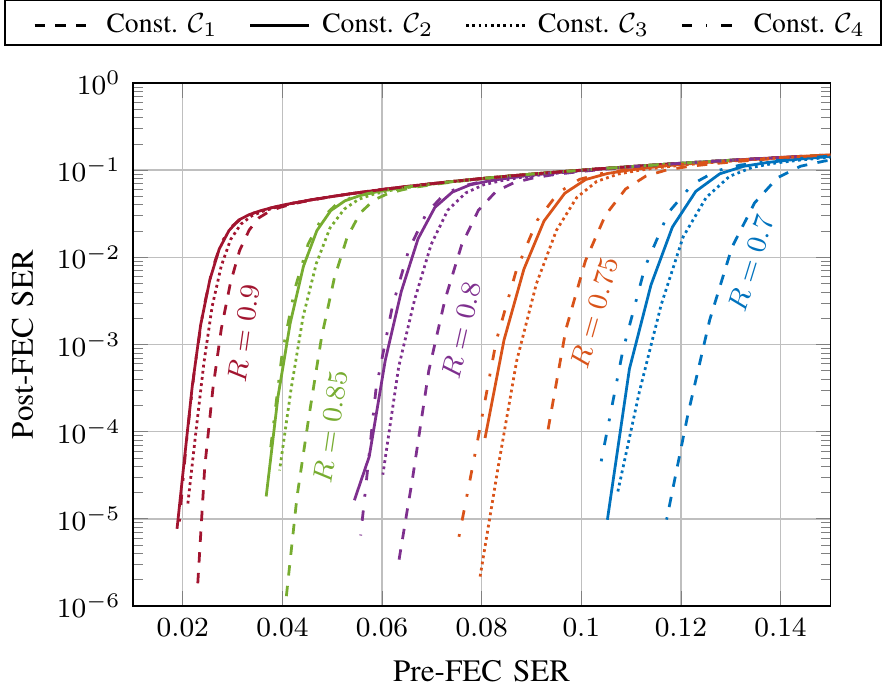}
\vspace*{-4ex}
\caption{Post-FEC SER as a function of the pre-FEC SER for the five LDPC codes of Tab.~\ref{tab:MIthresholds} using the four constellations of Fig.~\ref{fig:constellations} after transmission over an AWGN channel with hard symbol decision at the output.}\label{fig:HDLDPC_SER}
\end{figure}

\begin{figure}[tb!]
\centering
\includegraphics{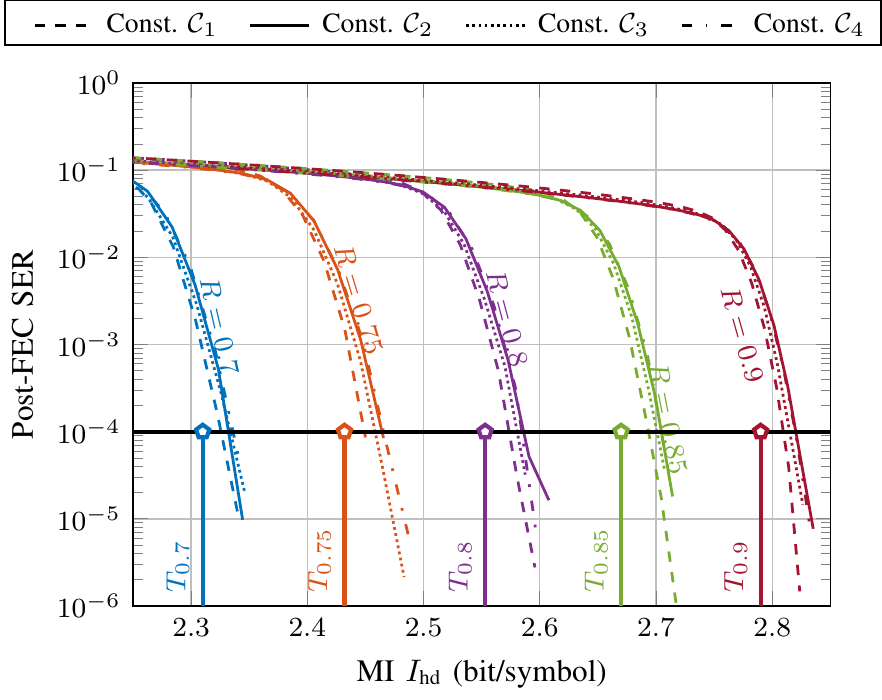}
\vspace*{-4ex}
\caption{Post-FEC SER as a function of the hard-decision MI $I_{\text{hd}}$ for the five LDPC codes of Tab.~\ref{tab:MIthresholds} using the four constellations of Fig.~\ref{fig:constellations} after transmission over an AWGN channel with hard symbol decision at the output.}\label{fig:HDLDPC_MI}
\end{figure}

Figures~\ref{fig:HDLDPC_SER} and~\ref{fig:HDLDPC_MI} show the post-FEC SER as a function of the pre-FEC SER after 15 LDPC decoding iterations with exactly the same decoder setup as used in Fig.~\ref{fig:comparisonMI}. We can clearly see that the pre-FEC SER is again not a good performance indicator while the MI is. For comparison, we also plot in Fig.~\ref{fig:HDLDPC_MI}	 the \ac{MI} thresholds for the different codes from Tab.~\ref{tab:MIthresholds}. We can see that the thresholds are not as precise as previously but still reflect the actual decoding performance. We attribute this offset to the fact that the utilized \ac{LDPC} codes  are not exactly universal and the length of the codes is relatively small, which is an effect that has also been observed in~\cite{Franceschini06}. Furthermore, we use off-the-shelf \ac{NB}-\ac{LDPC} codes with regular, unoptimized degree distributions. If we are allowed to increase the length of the codes and optimize the degree distributions, as highlighted for instance in~\cite{sanaei2008design}, the performance prediction becomes more accurate again. 

We hence conclude that the \ac{MI} is still an accurate estimate of the \ac{NB}-\ac{FEC} decoding performance, even if we introduce drastic changes into the channel (like, e.g., a hard decision, going from dispersion uncompensated to dispersion compensated link, or even from coherent transmission to direct detection systems). We can improve the accuracy if the channel that is used to compute the threshold is fairly close to the channel of the system.

\section{Conclusions}
Different performance metrics for coded modulation based on capacity-approaching \emph{nonbinary} codes were compared. It was shown in simulations and experiments that an accurate predictor of the performance of these codes is the mutual information, even under severe changes of the channel. Uncoded metrics such as pre-FEC~\ac{BER} and pre-FEC~\ac{SER} were shown to fail. The \ac{GMI} also fails for nonbinary codes, but still remains a good performance indicator for \ac{BICM} with \emph{binary} soft-decision \ac{FEC}. We have further discussed that it is necessary that the utilized codes are universal, which is however the case for most popular \ac{FEC} schemes used in optical communications.

\section*{Acknowledgment}
Laurent Schmalen would like to thank Dr. Georg B\"ocherer from Technical University Munich for stimulating discussions regarding mismatched decoding.



\begin{thebibliography}{10}
\providecommand{\url}[1]{#1}
\csname url@samestyle\endcsname
\providecommand{\newblock}{\relax}
\providecommand{\bibinfo}[2]{#2}
\providecommand{\BIBentrySTDinterwordspacing}{\spaceskip=0pt\relax}
\providecommand{\BIBentryALTinterwordstretchfactor}{4}
\providecommand{\BIBentryALTinterwordspacing}{\spaceskip=\fontdimen2\font plus
\BIBentryALTinterwordstretchfactor\fontdimen3\font minus
  \fontdimen4\font\relax}
\providecommand{\BIBforeignlanguage}[2]{{%
\expandafter\ifx\csname l@#1\endcsname\relax
\typeout{** WARNING: IEEEtran.bst: No hyphenation pattern has been}%
\typeout{** loaded for the language `#1'. Using the pattern for}%
\typeout{** the default language instead.}%
\else
\language=\csname l@#1\endcsname
\fi
#2}}
\providecommand{\BIBdecl}{\relax}
\BIBdecl

\bibitem{schmalen_ofc16}
L.~Schmalen, A.~Alvarado, and R.~Rios-M\"{u}ller, ``Predicting the performance
  of nonbinary forward error correction in optical transmission experiments,''
  in \emph{Optical Fiber Communication Conference (OFC)}, 2016, p. M2A.2.

\bibitem{Beppu15}
S.~Beppu, K.~Kasai, M.~Yoshida, and M.~Nakazawa, ``{2048 QAM (66 Gbit/s)}
  single-carrier coherent optical transmission over 150 km with a potential
  {SE} of {15.3 bit/s/Hz},'' \emph{Opt. Express}, vol.~23, no.~4, pp.
  4960--4969, Feb. 2015.

\bibitem{Qian13ecoc}
D.~Qian, E.~Ip, M.-F. Huang, M.-J. Li, and T.~Wang, ``{698.5-Gb/s PDM-2048QAM}
  transmission over 3km multicore fiber,'' in \emph{Proc. European Conference
  on Optical Communication (ECOC)}, London, UK, Sep. 2013, p. Th.1.C.5.

\bibitem{Alvarado2015b_JLT}
A.~Alvarado, E.~Agrell, D.~Lavery, R.~Maher, and P.~Bayvel, ``Replacing the
  soft-decision {FEC} limit paradigm in the design of optical communication
  systems,'' \emph{J. Lightw. Technol.}, vol.~33, no.~20, pp. 4338--4352, Oct.
  2015, (Invited Paper).

\bibitem{Alvarado16a}
------, ``Corrections to `{Replacing the Soft-Decision {FEC} Limit Paradigm in
  the Design of Optical Communication Systems}','' \emph{J. Lightw. Technol.},
  vol.~34, no.~2, p. 722, Jan. 2016.

\bibitem{Fabregas08_Book}
A.~{Guill\'en i F\`abregas}, A.~Martinez, and G.~Caire, ``Bit-interleaved coded
  modulation,'' \emph{Foundations and Trends in Communications and Information
  Theory}, vol.~5, no. 1--2, pp. 1--153, 2008.

\bibitem{Alvarado15_Book}
L.~Szczecinski and A.~Alvarado, \emph{Bit-Interleaved Coded Modulation:
  Fundamentals, Analysis and Design}.\hskip 1em plus 0.5em minus 0.4em\relax
  Chichester, UK: John Wiley \& Sons, 2015.

\bibitem{Bulow2011b}
H.~B\"{u}low, {\"{U}. Abay}, A.~Schenk, and J.~B. Huber, ``Coded modulation of
  polarization- and space-multiplexed signals,'' in \emph{Proc. Asia
  Communications and Photonics Conference and Exhibition (ACP)}, Shanghai,
  China, Nov. 2011.

\bibitem{Alvarado2015_JLT}
A.~Alvarado and E.~Agrell, ``Four-dimensional coded modulation with bit-wise
  decoders for future optical communications,'' \emph{J. Lightw. Technol.},
  vol.~33, no.~10, pp. 1993--2003, May 2015.

\bibitem{DjordjevicNB}
I.~B. Djordjevic and B.~Vasic, ``Nonbinary {LDPC} codes for optical
  communication systems,'' \emph{{IEEE} Photon. Technol. Lett.}, vol.~17,
  no.~10, pp. 2224--2226, 2005.

\bibitem{BeygiCM}
L.~Beygi, E.~Agrell, J.~M. Kahn, and M.~Karlsson, ``Coded modulation for
  fiber-optic networks: Toward better tradeoff between signal processing
  complexity and optical transparent reach,'' \emph{{IEEE} Signal Process.
  Mag.}, vol.~31, no.~2, pp. 93--103, 2014.

\bibitem{wachsmann1999multilevel}
U.~Wachsmann, R.~F.~H. Fischer, and J.~B. Huber, ``Multilevel codes:
  theoretical concepts and practical design rules,'' \emph{{IEEE} Trans. Inf.
  Theory}, vol.~45, no.~5, pp. 1361--1391, 1999.

\bibitem{Bisplinghoff:16}
A.~Bisplinghoff, N.~Beck, M.~Ene, M.~Danninger, and T.~Kupfer, ``Phase slip
  tolerant, low power multi-level coding for {64QAM} with {12.9 dB NCG},'' in
  \emph{Optical Fiber Communication Conference}.\hskip 1em plus 0.5em minus
  0.4em\relax Optical Society of America, 2016, p. M3A.2.

\bibitem{montorsi2012analog}
G.~Montorsi, ``Analog digital belief propagation,'' \emph{{IEEE} Commun.
  Lett.}, vol.~16, no.~7, pp. 1106--1109, Jul. 2012.

\bibitem{awais2014vlsi}
M.~Awais, G.~Masera, M.~Martina, and G.~Montorsi, ``{VLSI} implementation of a
  non-binary decoder based on the analog digital belief propagation,''
  \emph{{IEEE} Trans. Signal Process.}, vol.~62, no.~15, pp. 3965--3975, Aug.
  2014.

\bibitem{beermann2015gpu}
M.~Beermann, E.~Monz{\'o}, L.~Schmalen, and P.~Vary, ``{GPU} accelerated belief
  propagation decoding of non-binary {LDPC} codes with parallel and sequential
  scheduling,'' \emph{Journal of Signal Processing Systems}, vol.~78, no.~1,
  pp. 21--34, 2015.

\bibitem{Shannon48}
C.~E. Shannon, ``A mathematical theory of communication,'' \emph{Bell System
  Technical Journal}, vol.~27, pp. 379--423 and 623--656, July and Oct. 1948.

\bibitem{LevenMI}
A.~Leven, F.~Vacondio, L.~Schmalen, S.~{ten Brink}, and W.~Idler, ``Estimation
  of soft {FEC} performance in optical transmission experiments,'' \emph{{IEEE}
  Photon. Technol. Lett.}, vol.~20, no.~23, pp. 1547--1549, 2011.

\bibitem{Brueninghaus05}
K.~Brueninghaus, D.~Ast\'{e}ly, T.~S\"{a}lzer, S.~Visuri, A.~Alexiou,
  S.~Karger, and G.-A. Seraji, ``Link performance models for system level
  simulations of broadband radio access systems,'' in \emph{Proc. IEEE
  International Symposium on Personal, Indoor and Mobile Communications
  (PIMRC)}, Berlin, Germany, Sep. 2006.

\bibitem{Wan06}
L.~Wan, S.~Tsai, and M.~Almgren, ``A fading-insensitive performance metric for
  a unified link quality model,'' in \emph{Proc. IEEE Wireless Communications
  and Networking Conference (WCNC)}, Las Vegas, NV, Apr. 2006.

\bibitem{Agrell2009_JLT}
E.~Agrell and M.~Karlsson, ``Power-efficient modulation formats in coherent
  transmission systems,'' \emph{J. Lightw. Technol.}, vol.~27, no.~22, pp.
  5115--5126, Nov. 2009.

\bibitem{Eriksson_MI_JLT}
T.~A. Eriksson, T.~Fehenberger, P.~A. Andrekson, M.~Karlsson, N.~Hanik, and
  E.~Agrell, ``Impact of {4D} channel distribution on the achievable rates in
  coherent optical communication experiments,'' \emph{J. Lightw. Technol.},
  vol.~34, no.~9, pp. 2256--2266, May 2016.

\bibitem{ErikssonOFC14}
T.~A. Eriksson, P.~Johannisson, E.~Agrell, P.~A. Andrekson, and M.~Karlsson,
  ``Biorthogonal modulation in 8 dimensions experimentally implemented as
  {2PPM-PS-QPSK},'' in \emph{Proc. Optical Fiber Communication Conference
  (OFC)}, San Francisco, CA, Mar. 2014.

\bibitem{Koike-AkinoECOC13}
T.~Koike-Akino, D.~S. Millar, K.~Kojima, and K.~Parsons, ``Eight-dimensional
  modulation for coherent optical communications,'' in \emph{Proc. European
  Conference on Optical Communication (ECOC)}, London, UK, Sep. 2013.

\bibitem{buchali_jlt15}
F.~Buchali, F.~Steiner, G.~B\"ocherer, L.~Schmalen, P.~Schulte, and W.~Idler,
  ``Rate adaptation and reach increase by probabilistically shaped {64-QAM}: an
  experimental demonstration,'' \emph{J. Lightw. Technol.}, vol.~34, no.~7, pp.
  1599--1609, Apr. 2016.

\bibitem{bocherer_bandwidth_2015}
G.~B{\"o}cherer, F.~Steiner, and P.~Schulte, ``Bandwidth efficient and
  rate-matched low-density parity-check coded modulation,'' \emph{{IEEE} Trans.
  Commun.}, vol.~63, no.~12, pp. 4651--4665, Dec. 2015.

\bibitem{FehenbergerPTL16}
T.~Fehenberger, R.~Maher, A.~Alvarado, P.~Bayvel, and N.~Hanik, ``Sensitivity
  gains by mismatched probabilistic shaping for optical communication
  systems,'' \emph{{IEEE} Photon. Technol. Lett.}, vol.~28, no.~7, pp.
  786--789, Apr. 2016.

\bibitem{Agrell16RS}
E.~Agrell, A.~Alvarado, and F.~R. Kschischang, ``Implications of information
  theory in optical fibre communications,'' \emph{Philosophical Transactions
  A}, Feb. 2016, (Invited Paper).

\bibitem{ryan2009channel}
W.~Ryan and S.~Lin, \emph{Channel codes: classical and modern}.\hskip 1em plus
  0.5em minus 0.4em\relax Cambridge University Press, 2009.

\bibitem{merhav1994information}
N.~Merhav, G.~Kaplan, A.~Lapidoth, and S.~{Shamai (Shitz)}, ``On information
  rates for mismatched decoders,'' \emph{{IEEE} Trans. Inf. Theory}, vol.~40,
  no.~6, pp. 1953--1967, 1994.

\bibitem{poggiolini2012gn}
P.~Poggiolini, ``The {GN} model of non-linear propagation in uncompensated
  coherent optical systems,'' \emph{Journal of Lightwave Technology}, vol.~30,
  no.~24, pp. 3857--3879, 2012.

\bibitem{Franceschini06}
M.~Franceschini, G.~Ferrari, and R.~Raheli, ``Does the performance of {LDPC}
  codes depend on the channel?'' \emph{{IEEE} Trans. Commun.}, vol.~54, no.~12,
  pp. 2129--2132, Dec. 2006.

\bibitem{SasonUniversal}
I.~Sason and B.~Shuval, ``On universal {LDPC} code ensembles over memoryless
  symmetric channels,'' \emph{{IEEE} Trans. Inf. Theory}, vol.~57, no.~8, pp.
  5182--5202, Aug. 2011.

\bibitem{sanaei2008design}
A.~Sanaei, M.~Ramezani, and M.~Ardakani, ``On the design of universal {LDPC}
  codes,'' in \emph{Proc. IEEE ISIT}.\hskip 1em plus 0.5em minus 0.4em\relax
  IEEE, 2008, pp. 802--806.

\bibitem{SchmalenSCJLT}
L.~Schmalen, V.~Aref, J.~Cho, D.~Suikat, D.~R\"osener, and A.~Leven,
  ``Spatially coupled soft-decision error correction for future lightwave
  systems,'' \emph{J. Lightw. Technol.}, vol.~33, no.~5, pp. 1109--1116, Mar.
  2015.

\bibitem{Kudekarxx13}
S.~Kudekar, T.~Richardson, and R.~Urbanke, ``Spatially coupled ensembles
  universally achieve capacity under belief propagation,'' \emph{{IEEE} Trans.
  Inf. Theory}, vol.~59, no.~12, pp. 7761--7813, 2013.

\bibitem{ArikanPolar}
E.~{Ar\i kan}, ``Channel polarization: A method for constructing
  capacity-achieving codes for symmetric binary-input memoryless channels,''
  \emph{{IEEE} Trans. Inf. Theory}, vol.~55, no.~7, 2009.

\bibitem{verdu1994general}
S.~Verd\'u and T.~S. Han, ``A general formula for channel capacity,''
  \emph{{IEEE} Trans. Inf. Theory}, vol.~40, no.~4, pp. 1147--1157, 1994.

\bibitem{Agrell2014_JLT}
E.~Agrell, A.~Alvarado, G.~Durisi, and M.~Karlsson, ``Capacity of a nonlinear
  optical channel with finite memory,'' \emph{J. Lightw. Technol.}, vol.~32,
  no.~16, pp. 2862--2876, Aug. 2014 (Invited Paper).

\bibitem{Liga_submission16}
G.~Liga, A.~Alvarado, E.~Agrell, and P.~Bayvel, ``Information rates of
  next-generation long-haul optical fiber systems using coded modulation,''
  \emph{preprint available at arXiv.org}, 2016.

\bibitem{essiambre2010capacity}
R.-J. Essiambre, G.~Kramer, P.~J. Winzer, G.~J. Foschini, and B.~Goebel,
  ``Capacity limits of optical fiber networks,'' \emph{J. Lightw. Technol.},
  vol.~28, no.~4, pp. 662--701, 2010.

\bibitem{gantimismatched}
A.~Ganti, A.~Lapidoth, and {\.{I}. E}.~Telatar, ``Mismatched decoding
  revisited: General alphabets, channels with memory, and the wide-band
  limit,'' \emph{{IEEE} Trans. Inf. Theory}, vol.~46, no.~7, pp. 2315--2328,
  Nov. 2000.

\bibitem{GoldenMethod}
W.~Cheney and D.~Kincaid, \emph{Numerical Mathematics and Computing},
  3rd~ed.\hskip 1em plus 0.5em minus 0.4em\relax Brooks/Cole Publishing
  Company, 1994.

\bibitem{CoverThomas}
T.~M. Cover and J.~A. Thomas, \emph{Elements of Information Theory},
  2nd~ed.\hskip 1em plus 0.5em minus 0.4em\relax Wiley-Interscience, 2006.

\bibitem{yankov_constellation_2014}
M.~Yankov, D.~Zibar, K.~Larsen, L.~Christensen, and S.~Forchhammer,
  ``Constellation shaping for fiber-optic channels with {QAM} and high spectral
  efficiency,'' \emph{{IEEE} Photon. Technol. Lett.}, vol.~26, no.~23, pp.
  2407--2410, Dec. 2014.

\bibitem{richardsonmodern}
T.~Richardson and R.~Urbanke, \emph{Modern Coding Theory}.\hskip 1em plus 0.5em
  minus 0.4em\relax Cambridge Univ. Press, 2008.

\bibitem{silverman1986density}
B.~W. Silverman, \emph{Density Estimation for Statistics and Data
  Analysis}.\hskip 1em plus 0.5em minus 0.4em\relax CRC press, 1986, vol.~26.

\bibitem{arnold2006simulation}
D.~M. Arnold, H.-A. Loeliger, P.~O. Vontobel, A.~Kav{\v{c}}i{\'c}, and W.~Zeng,
  ``Simulation-based computation of information rates for channels with
  memory,'' \emph{{IEEE} Trans. Inf. Theory}, vol.~52, no.~8, pp. 3498--3508,
  2006.

\bibitem{fehenberger2015achievable}
T.~Fehenberger, A.~Alvarado, P.~Bayvel, and N.~Hanik, ``On achievable rates for
  long-haul fiber-optic communications,'' \emph{Optics Express}, vol.~23,
  no.~7, pp. 9183--9191, 2015.

\bibitem{secondini2013achievable}
M.~Secondini, E.~Forestieri, and G.~Prati, ``Achievable information rate in
  nonlinear {WDM} fiber-optic systems with arbitrary modulation formats and
  dispersion maps,'' \emph{J. Lightw. Technol.}, vol.~31, no.~23, pp.
  3839--3852, 2013.

\bibitem{feldman2005}
J.~Feldman, M.~J. Wainwright, and D.~R. Karger, ``Using linear programming to
  decode binary linear codes,'' \emph{{IEEE} Trans. Inf. Theory}, vol.~51,
  no.~3, pp. 954--972, Mar. 2005.

\bibitem{Li97}
X.~Li and J.~A. Ritcey, ``Bit-interleaved coded modulation with iterative
  decoding,'' \emph{{IEEE} Commun. Lett.}, vol.~1, no.~6, pp. 169--171, Nov.
  1997.

\bibitem{Brink98}
S.~{ten Brink}, J.~Speidel, and R.-H. Yan, ``Iterative demapping for {QPSK}
  modulation,'' \emph{IEE Electronics Letters}, vol.~34, no.~15, pp.
  1459--1460, July 1998.

\bibitem{Djordjevic2007_JLT}
I.~B. Djordjevic, M.~Cvijetic, L.~Xu, and T.~Wang, ``Using {LDPC}-coded
  modulation and coherent detection for ultra highspeed optical transmission,''
  \emph{J. Lightw. Technol.}, vol.~25, no.~11, pp. 3619--3625, Nov. 2007.

\bibitem{Batshon2009_JLT}
H.~B. Batshon, I.~B. Djordjevic, L.~Xu, and T.~Wang, ``Multidimensional
  {LDPC}-coded modulation for beyond {400 Gb/s} per wavelength transmission,''
  \emph{{IEEE} Photon. Technol. Lett.}, vol.~21, no.~16, pp. 1139--1141, Aug.
  2009.

\bibitem{Bulow14}
H.~B\"{u}low, X.~Lu, L.~Schmalen, A.~Klekamp, and F.~Buchali, ``Experimental
  performance of {4D} optimized constellation alternatives for {PM-8QAM} and
  {PM-16QAM},'' in \emph{Proc. Optical Fiber Communication Conference (OFC)},
  San Francisco, CA, Mar. 2014.

\bibitem{Bulow2011}
H.~B\"{u}low and E.~Masalkina, ``Coded modulation in optical communications,''
  in \emph{Proc. Optical Fiber Communication Conference (OFC)}, Los Angeles,
  CA, Mar. 2011.

\bibitem{Schmalen14}
L.~Schmalen, ``Energy efficient {FEC} for optical transmission systems,'' in
  \emph{Proc. Optical Fiber Communication Conference (OFC)}, San Francisco, CA,
  Mar. 2014.

\bibitem{SchmalenOFC15}
L.~Schmalen, S.~{ten Brink}, and A.~Leven, ``Spatially-coupled {LDPC}
  protograph codes for universal phase slip-tolerant differential decoding,''
  in \emph{Proc. Optical Fiber Communication Conference (OFC)}.\hskip 1em plus
  0.5em minus 0.4em\relax Optical Society of America, Mar. 2015, pp. Th3E--6.

\bibitem{SchmalenAdvances}
------, ``Advances in detection and error correction for coherent optical
  communications: Regular, irregular, and spatially coupled {LDPC} code
  designs,'' in \emph{Enabling Technologies for High Spectral-Efficiency
  Coherent Optical Communication Networks}, X.~Zhou and C.~Xie, Eds.\hskip 1em
  plus 0.5em minus 0.4em\relax Hoboken, NJ, USA: John Wiley \& Sons, Inc., Mar.
  2016, pp. 65--122.

\bibitem{SchmalenOFC12}
L.~Schmalen and R.~Dischler, ``Experimental evaluation of coded modulation for
  a coherent {PDM} system with high spectral efficiency,'' in \emph{Proc.
  Optical Fiber Communication Conference (OFC)}, 2012, pp. OW1H--1.

\bibitem{SchmalenSCC13}
L.~Schmalen and S.~{ten Brink}, ``Combining spatially coupled {LDPC} codes with
  modulation and detection,'' in \emph{Proc. International {ITG} Conference on
  Systems, Communication and Coding (SCC)}, Munich, Germany, Jan. 2013.

\bibitem{RiosMullerQAM8}
R.~Rios-M\"{u}ller, J.~Renaudier, L.~Schmalen, and G.~Charlet, ``Joint coding
  rate and modulation format optimization for {8QAM} constellations using
  {BICM} mutual information,'' in \emph{Optical Fiber Communication
  Conference}, 2015, p. W3K.4.

\bibitem{RiosMullerECOC14}
R.~Rios-M\"uller, J.~Renaudier, P.~Tran, and G.~Charlet, ``Experimental
  comparison of two {8-QAM} constellations at 200 {Gb/s} over ultra long-haul
  transmission link,'' in \emph{Proc. ECOC}, Cannes, France, Sep. 2014, p.
  P.5.1.

\bibitem{GellerNonbinary}
L.~Geller and D.~Burshtein, ``Bounds on the belief propagation threshold of
  non-binary {LDPC} codes,'' \emph{{IEEE} Trans. Inf. Theory}, vol.~62, no.~5,
  pp. 2639--2657, May 2016.

\bibitem{schmalen_generic_2012}
L.~Schmalen, F.~{Buchali}, A.~{Leven}, and S.~{ten Brink}, ``A generic tool for
  assessing the soft-{FEC} performance in optical transmission experiments,''
  \emph{{IEEE} Photon. Technol. Lett.}, vol.~24, no.~1, pp. 40--42, Jan. 2012.

\bibitem{stojanovic_reusing_2013}
N.~Stojanovic, Y.~{Zhao}, D.~{Chang}, Z.~{Xiao}, and F.~{Yu}, ``Reusing common
  uncoded experimental data in performance estimation of different {FEC}
  codes,'' \emph{{IEEE} Photon. Technol. Lett.}, vol.~25, no.~24, pp.
  2494--2497, Dec. 2013.

\end{thebibliography}
\end{document}